\documentclass[aip,reprint]{revtex4-1}

\usepackage{graphicx,amssymb,soul,color,amsmath,bm}
\usepackage{epstopdf,hyperref}
\usepackage{braket,dsfont,nicefrac}
\usepackage[mathcal]{euscript}
\usepackage[normalem]{ulem}
\usepackage[english]{babel}

\hypersetup{colorlinks=true,citecolor=blue,linkcolor=magenta}
\newcommand{\figwidth}{0.34\textwidth}
\graphicspath{{figs/}}

\draft 

\begin{document}


\title{A graphene edge-mediated quantum gate} 



\author{Phillip Weinberg}
\email[]{p.weinberg@northeastern.edu}
\affiliation{Department of Physics, Northeastern University, Boston, Massachusetts 02115, USA}

\author{Adrian E. Feiguin}
\affiliation{Department of Physics, Northeastern University, Boston, Massachusetts 02115, USA}

\date{\today}

\begin{abstract}
We propose a quantum gate architecture that allows for the systematic control of the effective exchange interactions between magnetic impurities embedded in nano-scale graphene flakes connected by a gated bridge.  The entanglement between the magnetic moment and the edge states of the fragments is used to electrostatically tune the exchange interaction from ferro to antiferromagnetic by merely changing the bridge's carrier density. By characterizing the effects of size and coupling parameters, we explore different operation regimes of this device by means of exact calculations with the density matrix renormalization group (DMRG). We analyze the results utilizing a simplified model that accounts for the main many-body mechanisms. Finally, we discuss how to use arrays of these devices to build quantum simulators for quantum many-body Hamiltonians.
\end{abstract}


\maketitle 

From a theoretical perspective, the exponential complexity of a quantum many-body wavefunction poses a significant challenge to our understanding of condensed matter physics both in and out of equilibrium. A practical and generic solution may lie in using nature itself as a quantum simulator~\cite{feynman86,lloyd96,trabesinger12,georgescu14}. The past few decades have witnessed remarkable progress towards a universal quantum simulator due to highly controllable experimental platforms such as ultracold gases \cite{jaksch05,garc_a_ripoll05,diehl08,bakr09,simon11,bloch12}, photonic devices \cite{aspuru-guzik12,peruzzo14,hartmann16,noh16,c_harris17},
polaritons \cite{berloff17}, trapped ions \cite{cirac95,james98,porras04,friedenauer08,haffner08,kim10,barreiro11}, Nitrogen vacancy centers~\cite{j_cai13}, and superconducting qubits~\cite{macha14,yan19,tan19,w_cai19,xu20,google20,r_harris10,r_harris10_2,dickson13,johnson11}. However, even with the multitude of experimental platforms that exist today, each one has its own unique limitation spanning from the types of interactions possible, to the available interaction graph. Most of these platforms are already in use as niche quantum simulators by taking advantage of their strengths, such as using D-wave devices to explore both equilibrium and non-equilibrium phenomena in quantum Ising models~\cite{harris18,gardas18,weinberg20,bando20}. 

The unique properties of graphene \cite{Novoselov.Electric_field_atomically_thin_carbon,graphene1b,geim2007} have made it very attractive as a potential replacement for silicon and as a platform to build quantum devices with diverse functionality, ranging from sensors to data storage \cite{Nalwa.MagneticNanostructures,avouris2007}.
In this context, a great deal of research exists on the physics of magnetic ad-atoms placed on mono-layer graphene sheets both experimentally and theoretically\cite{castro2009adatoms,Lars, shytov2009long,Kotov2012,graphene_review1,graphene_review2}. 
For magnetic impurities placed directly on top of carbon sites, theory predicts a Fermi liquid behavior consistent with an $SU(2)$ Kondo effect, also confirmed in experimental observations \cite{graphene_Kondo,cornaglia2009,sengupta2008,Jacob10}. 

Most work has focused on bulk --infinite-- graphene sheets, with a continuous density of states displaying a pseudo gap\cite{Uchoa2011}. However, the physics changes considerably if the impurity is placed in a finite graphene flake where the density of states consists of a discrete series of delta-like peaks, separated by a size-dependent energy spacing $\Delta$. In this case, the problem corresponds to that of a Kondo box \cite{Thimm1999,schlottmann2001kondo,simon2002finite,simon2003kondo,hand2006spin,Hanl2014}, Whether the impurity spin hybridizes with the conduction electrons or not depends on both $\Delta$ and the magnitude of the effective Kondo exchange $J$. It has been shown that the ``Kondo'' singlet in a finite graphene flake, is a ``few-electron problem'' \cite{Yang2017,debertolis2021fewbody}: If $J$ is small, the impurity will entangle primarily with state(s) near the Fermi energy; these states are delocalized plane-wave-like spreading through the entire volume of the substrate. However, if $J$ is large, the Kondo singlet will be localized in space and ``delocalized in energy''. Regardless of the case, the singlet state formed by the impurity and the conduction electrons will only involve one or very few electrons. A peculiarity about graphene, however, is that in a small flake, the state at the Fermi energy can correspond to an edge state. 

Typically edge states in graphene are exponentially localized near the edge\cite{Fujita1996}. However, in graphene flakes of irregular shape with a mixture of both zig-zag and arm chair edges, both theoretical and experimental studies \cite{Kobayashi2006} have shown edge states that can leak into the bulk. With this extended range, an impurity at the center of the flake can readily hybridize with the edge state of the flake. A small exchange $J$ will immediately entangle both, forming a correlated singlet that is practically decoupled from the bulk electrons\cite{Allerdt2017}.

\begin{figure}[t]
	\centering
	\includegraphics[width=3in]{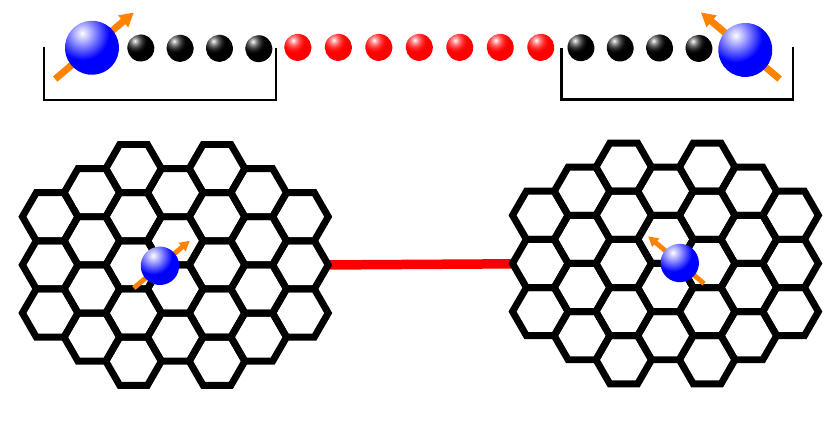}
	\caption{The top portion depicts the equivalent 1D model representing our proposed device in the calculations (see text). The blue sphere with the arrow represents the Kondo impurity, while the black sphere represent the electronic orbitals in the flake. The red spheres in the middle of the chain is the conducting bridge between the two flakes. The bottom panel shows a possible experimental implementation. A magnetic impurity sites at the center of graphene flakes connected by a 1D conducting channel.}
	\label{fig:bridge_fig}
\end{figure}

In the presence of two magnetic impurities, a new energy scale will arise, corresponding to the formation of a correlated Ruderman-Kittel-Kasuya-Yosida (RKKY)\cite{Yoisida.Magnetic_properties,Kittel.Indirect_exchange,Kasuya.Theory_of_metallic} singlet mediated by conduction electrons. We propose to use the edge electrons as the mediators of this effective exchange. Consider two flakes, such as illustrated in Fig.~\ref{fig:bridge_fig}, connected by a one-dimensional bridge. If the Kondo interaction is of the order of the energy splitting $\Delta$, the impurities will ``talk'' primarily to the edge electrons, which in turn hybridize with the electrons in the bridge. This will enable the impurities to form a correlated state across the bridge. By electrostatically tuning the position of the Fermi energy in the bridge, one can have exquisite control of the effective RKKY exchange and vary it from antiferromagnetic (AFM) to ferromagnetic (FM). 

In this letter, we study an idealized template that motivates further theoretical and experimental work on graphene devices like  the geometry depicted in Fig.~\ref{fig:bridge_fig}. In particular, we present an approximation of such a device and solve the many-body problem numerically; we discuss the phenomenology in terms of the entanglement structure of the exact many-body wavefunction; we conclude with a brief summary and outlook.\\

In order to construct a model approximating the geometry depicted in Fig.~\ref{fig:bridge_fig} we use a numerical approach developed to study the two impurity Kondo problems in monolayer graphene previously introduced in Ref.~\onlinecite{Allerdt2017}. That method is based on previous results showing how some impurity Hamiltonians can be mapped, via an exact unitary transformation onto a one dimensional (1D) or quasi-1D Hamiltonian\cite{Busser2013,Allerdt.Kondo,Frontiers_review}. We construct a model for the device by using the Hamiltonian for a finite-size graphene flake with radius $R$ and coupling two of these identical flakes together with a 1D channel modeled as a tight binding chain of length $L$ with nearest neighbor hopping. 

The model is pictorially illustrated in the upper part of Fig.~\ref{fig:bridge_fig}, the blue and cyan spheres on the left and right side of the chain (bracketed from below) represent the impurities and graphene flake orbitals respectively. The red spheres at the center represent the conducting channel orbitals. The resulting Hamiltonian is a 1D problem that is amenable to accurate numerical calculations, as we will discuss later. For the rest of this letter we work in units where $\hbar=1$ and the graphene hopping constant is $1$. The other adjustable parameters for our model are: the tight binding hopping parameter for the conducting channel, $t$; the hopping parameter between the graphene flakes and the conducting channel, $t'$; the gate voltage of the conducting channel, $V_g$; the Kondo coupling between graphene and the impurities, $J$. For the rest of this letter we will set $t=t'=1$. For more detailed description of this model see the supplemental material.

\begin{figure}[t]
	\includegraphics[width=\figwidth]{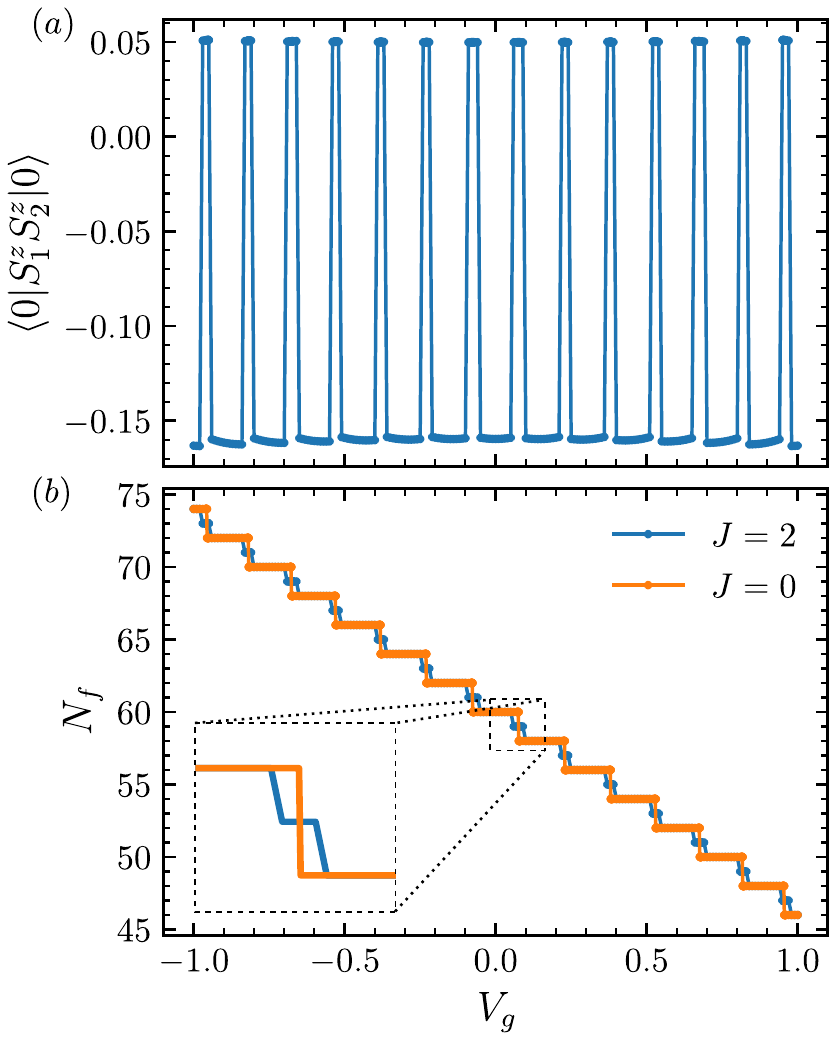}
	\caption{Panel $(a)$, the $S^z$ correlator between the two impurities as a function of the bridge gate potential, $V_g$ with $J=2$. Panel $(b)$, the electron occupation of the ground state as a function of $V_g$ for $J=2$ and $J=0$.}
	\label{fig:Nf_jumps}
\end{figure}

\begin{figure}[h]
	\includegraphics[width=\figwidth]{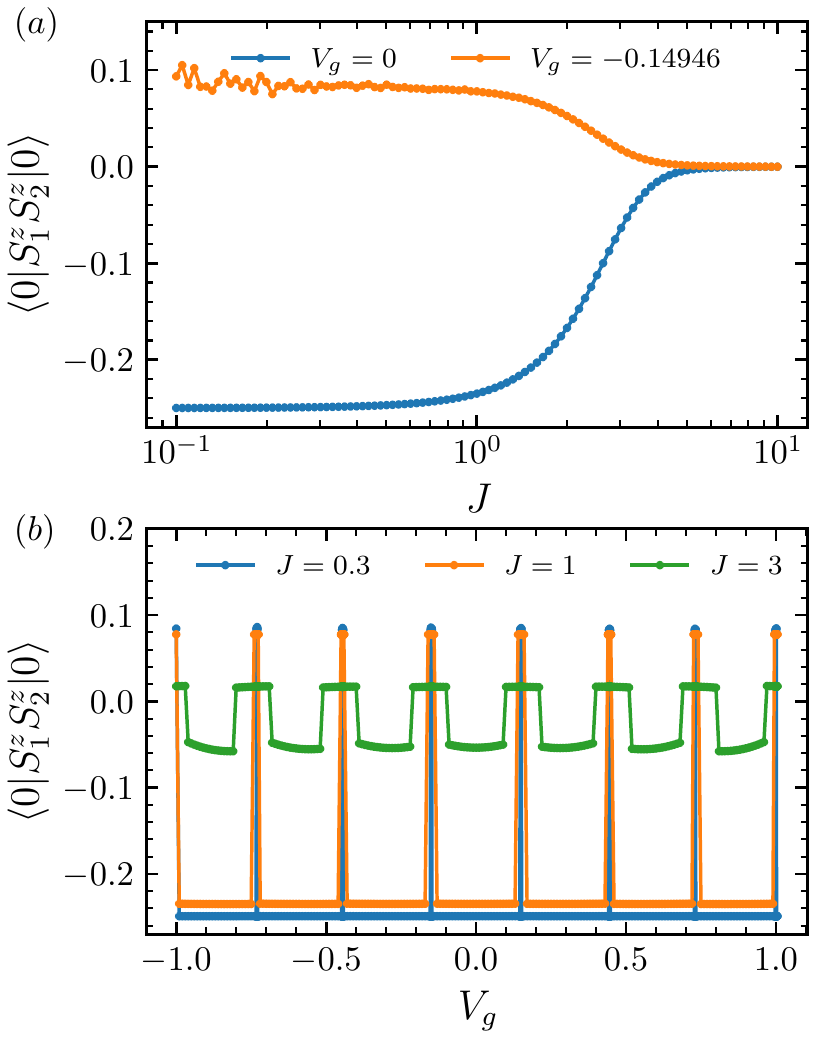}
	\caption{Panel $(a)$: $S^z$ correlator between the impurities for various values of the the Kondo coupling $J$ for two values of bridge gate potential such that each value corresponds to ferromagnetic correlations and antiferromagnetic correlations respectively. In the limit $J\rightarrow\infty$ the correlations decrease in magnitude indicating the formation of Kondo singlets. Panel $(b)$: The $S^z$ correlator between the two impurities for various values of the bridge chemical potential, $V_g$ with three different Kondo couplings $J$. For panel $(a)$ we fix $L=R=20$ and for panel $(b)$ we fix $L=20$ $R=10$.}
	\label{fig:corr_v_J}
\end{figure}

\begin{figure}[t!]
	\includegraphics[width=\figwidth]{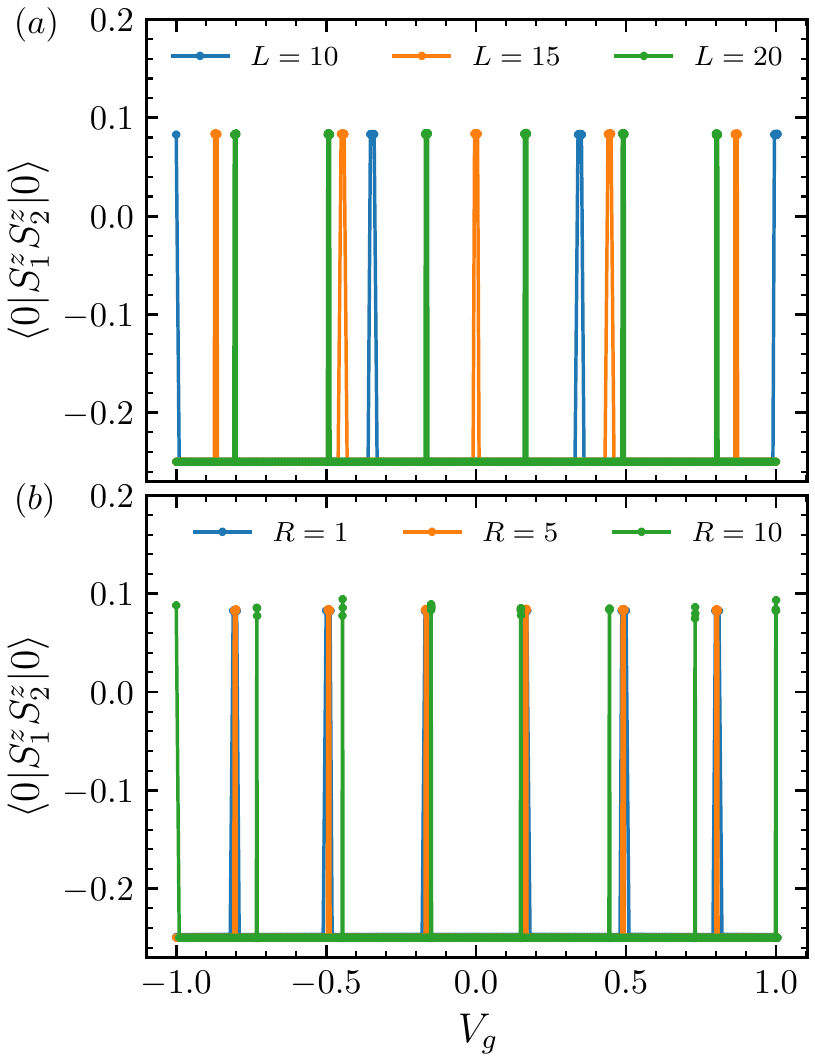}
	\caption{The $S^z$ correlator between the two impurities for various values of the bridge gate voltage, $V_g$ for: $(a)$ for different lengths of the bridge $L$, and $(b)$ different graphene flake radius $R$. For panel $(a)$ we fix $J=0.1$ and $R=5$ while for panel $(b)$ we fix $J=0.1$ and $L=20$.}
	\label{fig:LR_scaling}
\end{figure}

To study the ground state of our interacting Hamiltonian we use the density matrix renormalization group method (DMRG)\cite{White1992,White1993,Schollwock2005,Schollwock2011,Feiguin2013a} with fixed particle number. Once the ground state energies are obtained one can determine the total particle density as a function of $V_g$. In Fig.~\ref{fig:Nf_jumps} we present DMRG results with $R=10$, $L=40$, and $J=2$. We find there is switching between AFM to FM that correlates directly to jumps in the particle number. The positions of these jumps coincide with jumps in the non-interacting case (e.g. $J=0$), as shown in the figure. In the absence of interaction, the particle number changes by two when a discrete energy level crosses the Fermi energy. However, in the interacting case, the steps display an intermediate state with an odd particle number. The spin correlations are AFM for an even number of particles, to FM with odd.

Next, we discuss the effect of $J$ on the behavior of the correlations, and the competition between the formation of a correlated RKKY state or a screened Kondo singlet. In Fig.~\ref{fig:corr_v_J}$(a)$ we show the correlation between the impurities as a function of $J$. We observe that for a large range of $J$ the correlation remains $\mathcal{O}(1)$ even for non-perturbative values of the interaction. In Fig.~\ref{fig:corr_v_J}$(b)$ we plot the correlations vs. $V_g$ for three values of $J$. For larger $J$, the FM state is realized over a wider range albeit with a smaller magnitude. This effect indicates a means to stabilize the ferromagnetic spike-like behavior, which is a desired feature in real devices.

Finally, we study the correlation as a function of $L$. As $L$ increases, the energy spacing between single particle levels, $\Delta$, decreases. As such, we expect the number of jumps between AFM and FM correlations to increase which we observe in our simulations (see Fig.~\ref{fig:LR_scaling}$(a)$). On the other hand, the number of switching points does not change significantly with the radius of the flakes, as seen in Fig.~\ref{fig:LR_scaling}$(b)$. The numerical calculations provide compelling evidence for one of the main results of this work: the impurities are entangled to the zero-energy single particle level of the graphene flake, the edge state. As the radius increases, the amplitude of the edge state wavefunction at the impurity position will decrease, but this does not have a significant impact on the magnitude of the correlations. \\

\begin{figure}[t]
	\centering
	\includegraphics[width=3in]{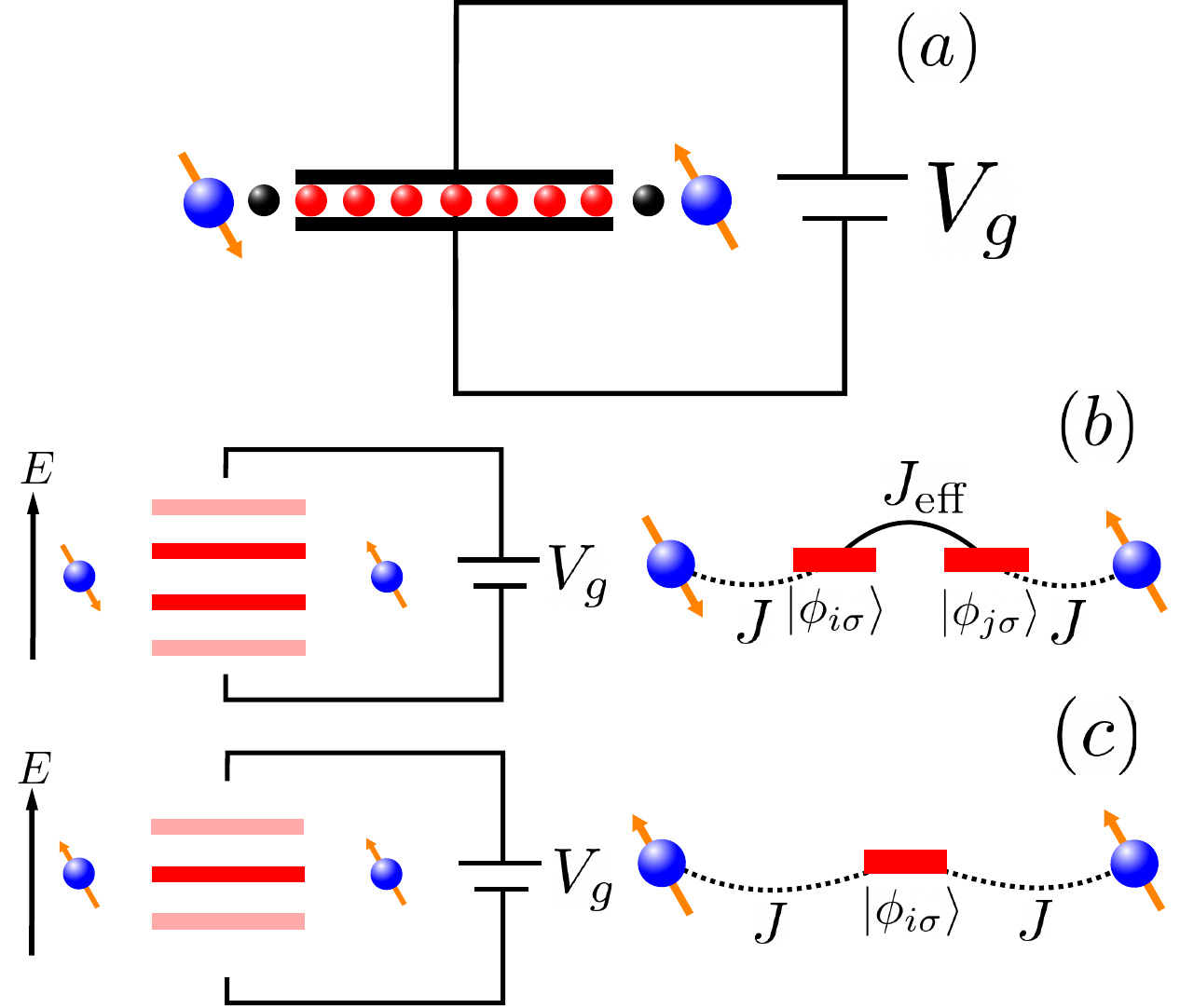}
	\caption{Label $(a)$ show a graphical representation of toy model used to understand the interaction between two Kondo impurities. The blue spheres with the arrows represent the Kondo impurities, the black spheres are the electronic orbitals that coupled the Kondo impurity to the rest of the system and the red spheres represent the bridge between the impurities that has chemical potential that is manipulated externally. Labels $(b)$ and $(c)$ depict the effective model in energy space and how that corresponds to an effective $4$ and $3$ site problem in the basis of natural orbitals respecitvely.}
	\label{fig:toy_model_fig}
\end{figure}

After establishing the behavior of the device, we can discuss the underlying mechanism using effective model amenable to exact numerical methods. As mentioned earlier, we assert that the edge states primarily mediate the RKKY impurity-impurity indirect exchange and that the bulk states of graphene do not play any role. Hence, it is reasonable to simply ``trace out'' the bulk states (see Fig.~\ref{fig:toy_model_fig}$(a)$ for a diagram). Along this line of reasoning, we consider our previous model Hamiltonian with $L=7$ and $R=1$ giving a total of $9$ electronic sites plus two impurities. Due to the small system size, we can use Lanczos to access the exact ground state wavefunction. In particular, we are interested in understanding the structure of the many-body wavefunction. 

\begin{figure}[t]
	\centering
	\includegraphics[width=\figwidth]{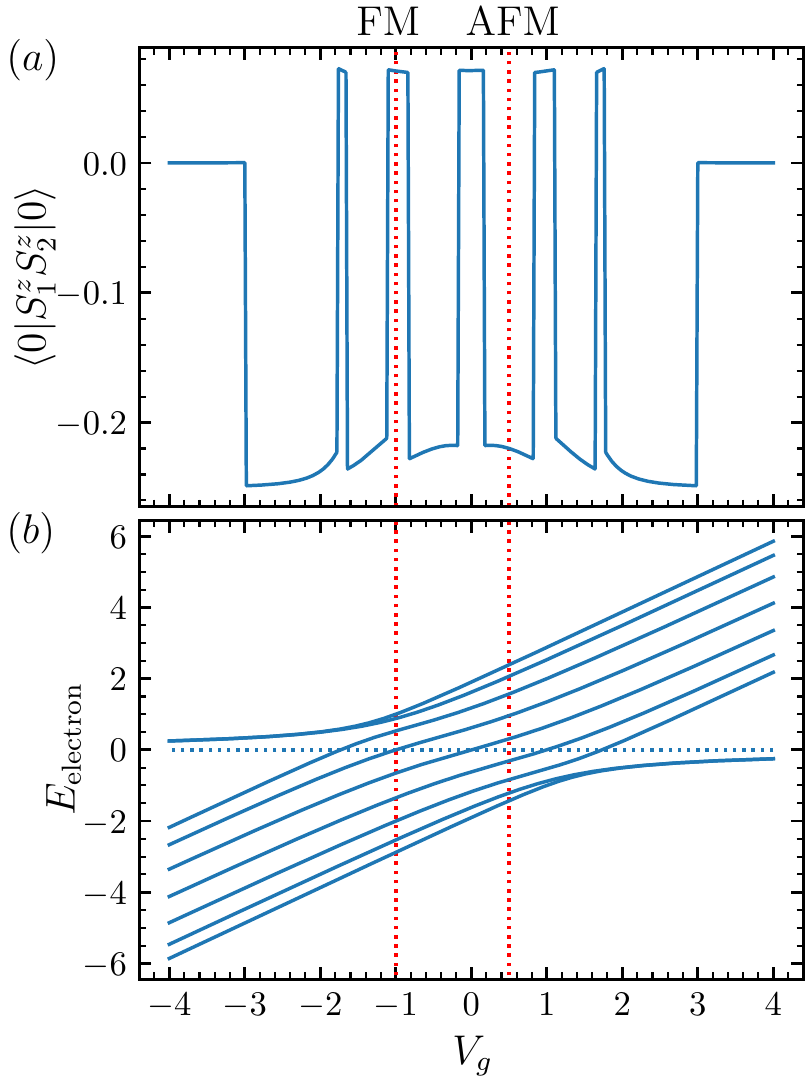}
	\caption{Panel $(a)$, the $S^z$ correlator between the two impurities as a function of the gate voltage on the bridge, $V_g$ with $J=0.3$. Panel $(b)$, the single-particle energies as a function of $V_g$. In both panels we fix $R=1$ and $L=7$. The vertical dotted lines are showing the position of specific values of $V_g$ we label with FM and AFM. These labels match the columns in Fig.~\ref{fig:slices_fig}. The horizontal dotted line in panel $(b)$ is there to emphasize where the lines cross a y-value of $0$.}
	\label{fig:correlation_fig}
\end{figure}

\begin{figure}[t]
	\centering
	\includegraphics[width=3.3in]{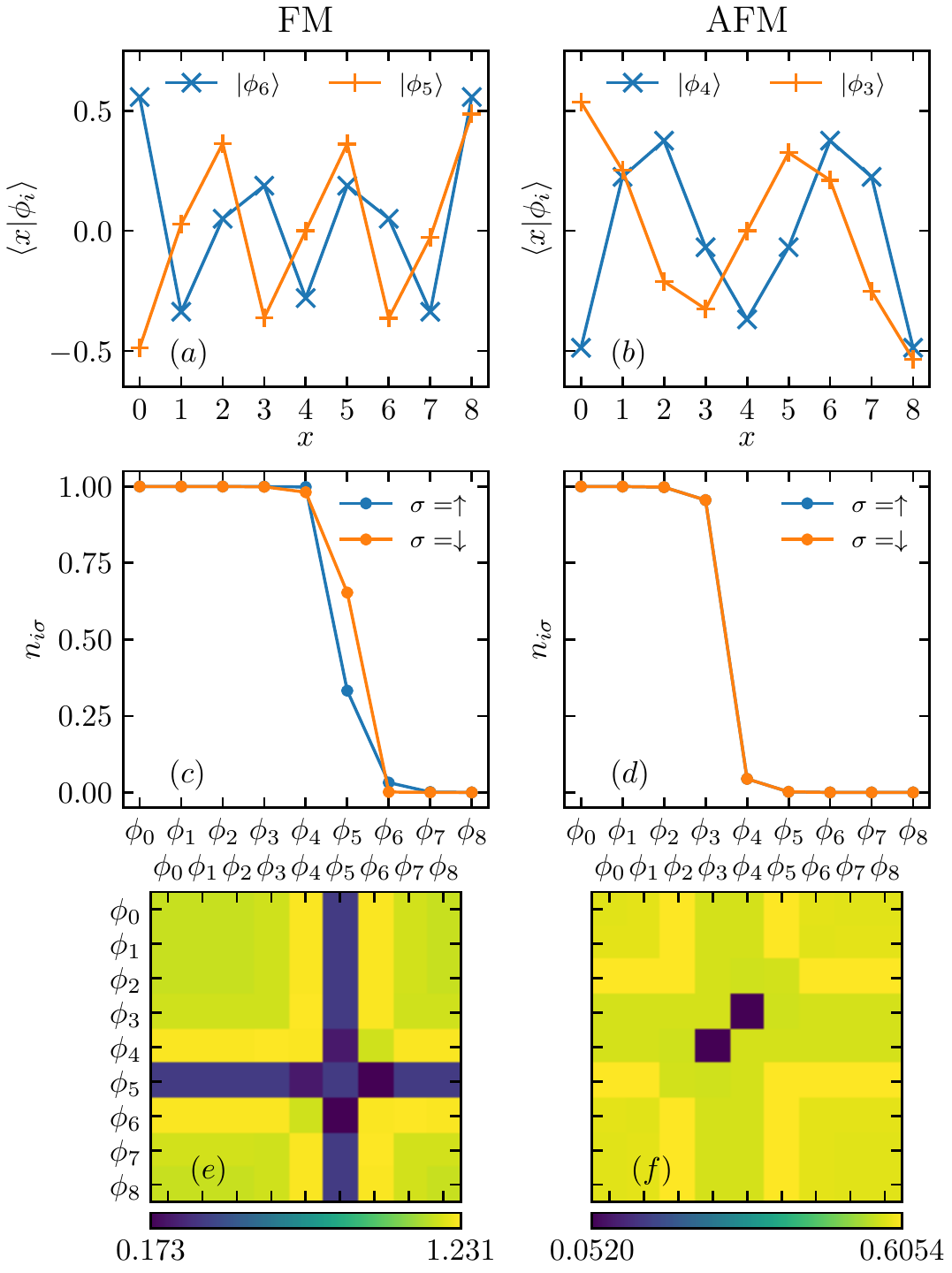}
	\caption{Collection of plots pertaining to the natural orbitals for the ground state of the graphene model with $R=1$ and $L=7$. Left and right columns correspond to the FM and AFM cases corresponding to values of $V_g$ labeled as $FM$ and $AFM$ denoted by vertical lines in Fig.~\ref{fig:correlation_fig}. $\phi_i$ corresponds to the $i$-th natural orbital. In each row we present: (top) the two orbitals that give the partition that minimize the entanglement projected to the local single-particle basis, (center) the occupations of the natural orbitals, and (bottom) the entanglement entropy for various partitions containing $1$ and $2$ natural orbitals and the two impurities as a heat map. }
	\label{fig:slices_fig}
\end{figure}

In Fig.~\ref{fig:correlation_fig}$(a)$ and $(b)$ we show the correlation between the two impurities and the single-particle spectrum respectively as a function of $V_g$ with $J=0.5$. In Fig.~\ref{fig:correlation_fig}$(b)$ we show the non-interacting single-particle electronic energies as a function of $V_g$. From that figure, we observe that the correlations are FM when there is a single-particle electronic state that has energy close to $E=0$ and AFM otherwise. The results mirror the behavior observed in the full model and further supports our argument that only states near the Fermi-energy contribute to the interactions between the impurities. In order to confirm this hypothesis definitively, we must delve into the structure of the many-body wavefunction. 

Assuming that one or two electrons are responsible for mediating the interaction between the impurities, it should be possible find one or two single-particle electronic wavefunctions that are entangled to the spins. The question then becomes: which single particle basis should we use to generate this partition. In previous work it was shown that the answer lies in using the natural orbitals of the system as the single particle basis\cite{Yang2017}. The natural orbitals are defined as the eigenstates of the single density matrix:
\begin{gather}
	G_{ij,\sigma} = \langle 0|c^\dagger_{\sigma i}c_{\sigma j}|0\rangle,\\
	\mathbf{G}_\sigma|\phi_{i,\sigma}\rangle = n_{i,\sigma}|\phi_{i,\sigma}\rangle.
\end{gather}
Here $c_{\sigma i}$ and $c^\dagger_{\sigma i}$  corresponds to the creation and annihilation operators of all the electronic orbitals respectively. From now on, we refer to the bi-partite entanglement in the basis of the natural orbitals meaning we will calculate the reduced density matrix of a partition that includes the impurities and one or two natural orbitals. 

We can determine when the subsystem effectively decouples from the rest of the many-body wavefunction by checking all possible partitions. In Fig.~\ref{fig:slices_fig}, we summarize this entanglement analysis for two values of $V_g$ denoted by vertical dashed red lines in Fig.~\ref{fig:correlation_fig}. The labels  FM and AFM in Fig.~\ref{fig:correlation_fig} corresponding the same labels in Fig.~\ref{fig:slices_fig}. The bottom row in Fig.~\ref{fig:slices_fig} contains density plots showing the entanglement entropy divided by $\log 2$ for every possible partition involving one and two natural orbitals. The diagonal part of the matrix denotes the partition with one orbital while the off diagonal elements denote partitions with two orbitals. Since the order of the orbitals in the partition does not change the entanglement entropy, this matrix is symmetric. The top row in Fig.~\ref{fig:slices_fig} shows the two natural orbitals that entangled to the impurities. The orbitals are shown in terms of the local single-particle basis. Finally, in the middle row of Fig.~\ref{fig:slices_fig} we show the occupation of the spin-up and spin-down natural orbitals. 

When the correlations are FM, the partition is minimized when entangled to only a single natural orbital. Furthermore, by looking at the occupation of this orbital, it is clear that only one electron at the fermi-energy is forming a global singlet with the two impurities. On the other hand, for AFM correlations, the entanglement and occupation imply that two electrons partially occupying two natural orbitals closest to the fermi-energy participate in the interaction between the impurities. 

From the results we conclude our intuition that the RKKY interaction between the impurities is mediated by, at most, two electrons such that the ground state wavefunction can be written, to a great degree of accuracy as:
\begin{eqnarray}
|\psi_{gs}\rangle \approx|\sigma\rangle \otimes |FS'\rangle,
\end{eqnarray}
where $|\sigma\rangle$ is a many body state involving the two impurity spins and the first and/or second natural orbitals, which is practically decoupled from a new Fermi sea $|FS'\rangle$ comprised by the remaining non-interacting electrons.

While the magnitude of the RKKY interactions will depend on the Kondo coupling and the position of the gate voltage, one can predict what the maximum value of the spin-spin correlations can be. For small Kondo coupling, the impurity spins are very weakly entangled, meaning that the effective RKKY interaction is vanishingly small. However, we observe that the spins are practically decoupled. In fact, we see that, since the total value of $S^z=0$, one of them has to point up, while the other down, which yields the observed value of the correlations $-1/4$. This corresponds to what is known as the free moment regime in which the Kondo coupling is not strong enough to create particle-hole excitations and entangle the impurity to the conduction electrons. As $J$ is increased, the impurities will for an RKKY singlet mediated by the natural orbitals as discussed in the previous sections. Hence, the fact that the correlations are practically saturated is not a sign of strong interactions, rather, it reflects the fact that their coupling to the conduction electrons is in fact very weak. At weak coupling, we observe that only two natural orbitals will contribute to the RKKY wave function. As $J$ increases, this number will also increase, and the value of the correlations will decrease. For the FM case, the gate voltage has to be such that there is always a single-particle state close to zero energy. Therefore, only one natural orbital will be participating. 

Therefore, an intuitive picture can be obtained by reducing the problem to an equivalent 4-site, or 3-site spin chain with antiferromagnetic interactions in order to describe the emerging physics (See Fig.~\ref{fig:toy_model_fig}$(b)$ and $(c)$ respectively). The interaction in the AFM case will have a stronger dependence on system size and the effective RKKY interaction, while for the FM case, the interaction will only depend on $J$. The value for the correlations in the 3-spin ground state is $1/12=0.08\overline{33}$, in agreement with the one observed numerically for the full problems.

\begin{figure}[t]
	\centering
	\includegraphics[width=3.5in]{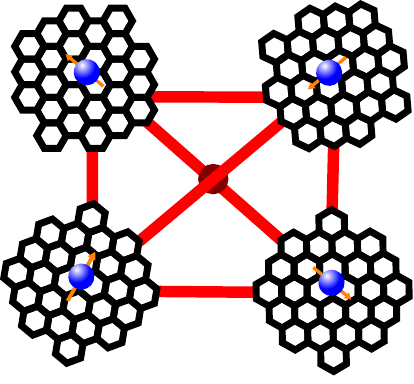}
	\caption{Schematic diagram illustrating how to build a simulator by scaling up the idea of a single quantum gate to an array multiple interconnected flakes. The bridges in red are back gated and in principle the can be tuned independently such that the sign of the interactions between impurity spins can be changed individually.}
	\label{fig:simulator}
\end{figure}

To conclude, we have presented exact numerical results for the ground state and correlation functions of a realistic model describing two magnetic ad-atoms at the center of graphene flakes coupled by a 1D conducting channel. While this arrangement, at first sight, resembles the well studied two-impurity Kondo problem, we find a peculiar behavior that stems from two identifiable properties of the system: (i) the flakes are each realization of the ``Kondo box'' with discrete energy levels and (ii) graphene flakes with mixed armchair and zig-zag edges exhibit zero-energy extended edge states that leak into the bulk. Consequently, the impurities will interact primarily with the electronic wavefunction of the edge states, which will mediate the effective RKKY exchange through the bridge. We have shown that the problem can be formally understood by ultimately tracing out the bulk states of graphene and only accounting for the edge electrons. In such a case, the exact geometry and size of the graphene flake do not play an important role. By gating the bridge, one can electrostatically tune the electronic density of the system and switch the RKKY interaction from ferro to antiferromagnetic with a great degree of control.

The two-qubit quantum gate that we present in this work would operate in a similar fashion as the two-qubit quantum gates on quantum dots, as discussed in a now ``classic'' paper by D. loss and D. DiVincenzo, Ref.~\onlinecite{divincenzo98}. However, we here can take advantage of the ``macroscopic'' nature of the entangled state between the spin and the edge state of the graphene flake, which could be several nanometers in size. In order to operate the gates, the spins will be subject to a transient coupling $J_{RKKY}(t)$ that would enable to perform the (square root of the) SWAP operator. To be able to carry out universal computations one would need operations such as the XOR gate, which in addition would require single qubit operations. In the aforementioned paper, the proposal consists of targeting the single spins with a scanning probe. While this is technologically possible for our proposed device, it would be challenging. Therefore, we suggest that the most straightforward and impactful application of this architecture would be to implement many-body interactions between localized spins/qubits in order to operate the device as a quantum simulator to probe for exotic magnetic quantum states of matter.

The above considerations enable us to propose this system as a building block for a graphene-based quantum simulator: one can envision an array of such graphene flakes connected by bridges forming a lattice (see Fig.~\ref{fig:simulator}. Back gates can tune the effective pair-wise exchange between magnetic ad-atoms from FM to AFM. By taking advantage of the entanglement between the impurities with the edge states, one does not need to address each impurity individually but electrostatically tune the density of the edge state. One can even vary the spin of individual impurities by using different transition metals, for instance.
As an example, porphyrin-like centers can be embedded in graphene and carbon nanotubes \cite{Zhang2009,Lee2011,Chung2013,Zhu2013,Jia2015,Tylus2016,Aoyama2018,Allerdt2020}
While the intrinsic form and anisotropy of the interaction between the impurity and graphene is difficult to control, one can access and manipulate the inter-impurity exchange over the entire array, offering the means to realize and study exotic quantum-many body states, such as quantum spin liquids. 

\section*{Supplemental Material}
As discussed in the main text, we refer the reader to the supplemental material for a more complete description of the graphene flake mapping as well as a brief discussion of the effect of varying $t'$ on the correlations between the impurities.  

\begin{acknowledgements}
The authors are supported by the US Department of Energy (DOE), Office of Science, Basic Energy Sciences grant number DE-SC0019275.
\end{acknowledgements}

\section*{Data Availability}
The code and any data generated by the code that support the findings of this study are available online at \url{https://github.com/weinbe58/graphene_gate_APL}

\bibliography{refs,Graphene_paper, impurities_graphene,dmrg,porphyrins}

\begin{thebibliography}{85}%
\makeatletter
\providecommand \@ifxundefined [1]{%
 \@ifx{#1\undefined}
}%
\providecommand \@ifnum [1]{%
 \ifnum #1\expandafter \@firstoftwo
 \else \expandafter \@secondoftwo
 \fi
}%
\providecommand \@ifx [1]{%
 \ifx #1\expandafter \@firstoftwo
 \else \expandafter \@secondoftwo
 \fi
}%
\providecommand \natexlab [1]{#1}%
\providecommand \enquote  [1]{``#1''}%
\providecommand \bibnamefont  [1]{#1}%
\providecommand \bibfnamefont [1]{#1}%
\providecommand \citenamefont [1]{#1}%
\providecommand \href@noop [0]{\@secondoftwo}%
\providecommand \href [0]{\begingroup \@sanitize@url \@href}%
\providecommand \@href[1]{\@@startlink{#1}\@@href}%
\providecommand \@@href[1]{\endgroup#1\@@endlink}%
\providecommand \@sanitize@url [0]{\catcode `\\12\catcode `\$12\catcode
  `\&12\catcode `\#12\catcode `\^12\catcode `\_12\catcode `\%12\relax}%
\providecommand \@@startlink[1]{}%
\providecommand \@@endlink[0]{}%
\providecommand \url  [0]{\begingroup\@sanitize@url \@url }%
\providecommand \@url [1]{\endgroup\@href {#1}{\urlprefix }}%
\providecommand \urlprefix  [0]{URL }%
\providecommand \Eprint [0]{\href }%
\providecommand \doibase [0]{http://dx.doi.org/}%
\providecommand \selectlanguage [0]{\@gobble}%
\providecommand \bibinfo  [0]{\@secondoftwo}%
\providecommand \bibfield  [0]{\@secondoftwo}%
\providecommand \translation [1]{[#1]}%
\providecommand \BibitemOpen [0]{}%
\providecommand \bibitemStop [0]{}%
\providecommand \bibitemNoStop [0]{.\EOS\space}%
\providecommand \EOS [0]{\spacefactor3000\relax}%
\providecommand \BibitemShut  [1]{\csname bibitem#1\endcsname}%
\let\auto@bib@innerbib\@empty
\bibitem [{\citenamefont {Feynman}(1986)}]{feynman86}%
  \BibitemOpen
  \bibfield  {author} {\bibinfo {author} {\bibfnamefont {R.~P.}\ \bibnamefont
  {Feynman}},\ }\bibfield  {title} {\enquote {\bibinfo {title} {Quantum
  mechanical computers},}\ }\href {\doibase 10.1007/BF01886518} {\bibfield
  {journal} {\bibinfo  {journal} {Foundations of Physics}\ }\textbf {\bibinfo
  {volume} {16}},\ \bibinfo {pages} {507--531} (\bibinfo {year}
  {1986})}\BibitemShut {NoStop}%
\bibitem [{\citenamefont {Lloyd}(1996)}]{lloyd96}%
  \BibitemOpen
  \bibfield  {author} {\bibinfo {author} {\bibfnamefont {S.}~\bibnamefont
  {Lloyd}},\ }\bibfield  {title} {\enquote {\bibinfo {title} {Universal quantum
  simulators},}\ }\href {http://www.jstor.org/stable/2899535} {\bibfield
  {journal} {\bibinfo  {journal} {Science}\ }\textbf {\bibinfo {volume}
  {273}},\ \bibinfo {pages} {1073--1078} (\bibinfo {year} {1996})},\ \bibinfo
  {note} {full publication date: Aug. 23, 1996}\BibitemShut {NoStop}%
\bibitem [{\citenamefont {Trabesinger}(2012)}]{trabesinger12}%
  \BibitemOpen
  \bibfield  {author} {\bibinfo {author} {\bibfnamefont {A.}~\bibnamefont
  {Trabesinger}},\ }\bibfield  {title} {\enquote {\bibinfo {title} {Quantum
  simulation},}\ }\href {\doibase 10.1038/nphys2258} {\bibfield  {journal}
  {\bibinfo  {journal} {Nature Physics}\ }\textbf {\bibinfo {volume} {8}},\
  \bibinfo {pages} {263--263} (\bibinfo {year} {2012})}\BibitemShut {NoStop}%
\bibitem [{\citenamefont {Georgescu}, \citenamefont {Ashhab},\ and\
  \citenamefont {Nori}(2014)}]{georgescu14}%
  \BibitemOpen
  \bibfield  {author} {\bibinfo {author} {\bibfnamefont {I.~M.}\ \bibnamefont
  {Georgescu}}, \bibinfo {author} {\bibfnamefont {S.}~\bibnamefont {Ashhab}}, \
  and\ \bibinfo {author} {\bibfnamefont {F.}~\bibnamefont {Nori}},\ }\bibfield
  {title} {\enquote {\bibinfo {title} {Quantum simulation},}\ }\href {\doibase
  10.1103/RevModPhys.86.153} {\bibfield  {journal} {\bibinfo  {journal} {Rev.
  Mod. Phys.}\ }\textbf {\bibinfo {volume} {86}},\ \bibinfo {pages} {153--185}
  (\bibinfo {year} {2014})}\BibitemShut {NoStop}%
\bibitem [{\citenamefont {Jaksch}\ and\ \citenamefont
  {Zoller}(2005)}]{jaksch05}%
  \BibitemOpen
  \bibfield  {author} {\bibinfo {author} {\bibfnamefont {D.}~\bibnamefont
  {Jaksch}}\ and\ \bibinfo {author} {\bibfnamefont {P.}~\bibnamefont
  {Zoller}},\ }\bibfield  {title} {\enquote {\bibinfo {title} {The cold atom
  hubbard toolbox},}\ }\href {\doibase
  https://doi.org/10.1016/j.aop.2004.09.010} {\bibfield  {journal} {\bibinfo
  {journal} {Annals of Physics}\ }\textbf {\bibinfo {volume} {315}},\ \bibinfo
  {pages} {52--79} (\bibinfo {year} {2005})},\ \bibinfo {note} {special
  Issue}\BibitemShut {NoStop}%
\bibitem [{\citenamefont {Garc{\'{\i}}a-Ripoll}, \citenamefont {Zoller},\ and\
  \citenamefont {Cirac}(2005)}]{garc_a_ripoll05}%
  \BibitemOpen
  \bibfield  {author} {\bibinfo {author} {\bibfnamefont {J.~J.}\ \bibnamefont
  {Garc{\'{\i}}a-Ripoll}}, \bibinfo {author} {\bibfnamefont {P.}~\bibnamefont
  {Zoller}}, \ and\ \bibinfo {author} {\bibfnamefont {J.~I.}\ \bibnamefont
  {Cirac}},\ }\bibfield  {title} {\enquote {\bibinfo {title} {Quantum
  information processing with cold atoms and trapped ions},}\ }\href {\doibase
  10.1088/0953-4075/38/9/008} {\bibfield  {journal} {\bibinfo  {journal}
  {Journal of Physics B: Atomic, Molecular and Optical Physics}\ }\textbf
  {\bibinfo {volume} {38}},\ \bibinfo {pages} {S567--S578} (\bibinfo {year}
  {2005})}\BibitemShut {NoStop}%
\bibitem [{\citenamefont {Diehl}\ \emph {et~al.}(2008)\citenamefont {Diehl},
  \citenamefont {Micheli}, \citenamefont {Kantian}, \citenamefont {Kraus},
  \citenamefont {B{\"u}chler},\ and\ \citenamefont {Zoller}}]{diehl08}%
  \BibitemOpen
  \bibfield  {author} {\bibinfo {author} {\bibfnamefont {S.}~\bibnamefont
  {Diehl}}, \bibinfo {author} {\bibfnamefont {A.}~\bibnamefont {Micheli}},
  \bibinfo {author} {\bibfnamefont {A.}~\bibnamefont {Kantian}}, \bibinfo
  {author} {\bibfnamefont {B.}~\bibnamefont {Kraus}}, \bibinfo {author}
  {\bibfnamefont {H.~P.}\ \bibnamefont {B{\"u}chler}}, \ and\ \bibinfo {author}
  {\bibfnamefont {P.}~\bibnamefont {Zoller}},\ }\bibfield  {title} {\enquote
  {\bibinfo {title} {Quantum states and phases in driven open quantum systems
  with cold atoms},}\ }\href {\doibase 10.1038/nphys1073} {\bibfield  {journal}
  {\bibinfo  {journal} {Nature Physics}\ }\textbf {\bibinfo {volume} {4}},\
  \bibinfo {pages} {878--883} (\bibinfo {year} {2008})}\BibitemShut {NoStop}%
\bibitem [{\citenamefont {Bakr}\ \emph {et~al.}(2009)\citenamefont {Bakr},
  \citenamefont {Gillen}, \citenamefont {Peng}, \citenamefont {F{\"o}lling},\
  and\ \citenamefont {Greiner}}]{bakr09}%
  \BibitemOpen
  \bibfield  {author} {\bibinfo {author} {\bibfnamefont {W.~S.}\ \bibnamefont
  {Bakr}}, \bibinfo {author} {\bibfnamefont {J.~I.}\ \bibnamefont {Gillen}},
  \bibinfo {author} {\bibfnamefont {A.}~\bibnamefont {Peng}}, \bibinfo {author}
  {\bibfnamefont {S.}~\bibnamefont {F{\"o}lling}}, \ and\ \bibinfo {author}
  {\bibfnamefont {M.}~\bibnamefont {Greiner}},\ }\bibfield  {title} {\enquote
  {\bibinfo {title} {A quantum gas microscope for detecting single atoms in a
  hubbard-regime optical lattice},}\ }\href {\doibase 10.1038/nature08482}
  {\bibfield  {journal} {\bibinfo  {journal} {Nature}\ }\textbf {\bibinfo
  {volume} {462}},\ \bibinfo {pages} {74--77} (\bibinfo {year}
  {2009})}\BibitemShut {NoStop}%
\bibitem [{\citenamefont {Simon}\ \emph {et~al.}(2011)\citenamefont {Simon},
  \citenamefont {Bakr}, \citenamefont {Ma}, \citenamefont {Tai}, \citenamefont
  {Preiss},\ and\ \citenamefont {Greiner}}]{simon11}%
  \BibitemOpen
  \bibfield  {author} {\bibinfo {author} {\bibfnamefont {J.}~\bibnamefont
  {Simon}}, \bibinfo {author} {\bibfnamefont {W.~S.}\ \bibnamefont {Bakr}},
  \bibinfo {author} {\bibfnamefont {R.}~\bibnamefont {Ma}}, \bibinfo {author}
  {\bibfnamefont {M.~E.}\ \bibnamefont {Tai}}, \bibinfo {author} {\bibfnamefont
  {P.~M.}\ \bibnamefont {Preiss}}, \ and\ \bibinfo {author} {\bibfnamefont
  {M.}~\bibnamefont {Greiner}},\ }\bibfield  {title} {\enquote {\bibinfo
  {title} {Quantum simulation of antiferromagnetic spin chains in an optical
  lattice},}\ }\href {\doibase 10.1038/nature09994} {\bibfield  {journal}
  {\bibinfo  {journal} {Nature}\ }\textbf {\bibinfo {volume} {472}},\ \bibinfo
  {pages} {307--312} (\bibinfo {year} {2011})}\BibitemShut {NoStop}%
\bibitem [{\citenamefont {Bloch}, \citenamefont {Dalibard},\ and\ \citenamefont
  {Nascimb{\`e}ne}(2012)}]{bloch12}%
  \BibitemOpen
  \bibfield  {author} {\bibinfo {author} {\bibfnamefont {I.}~\bibnamefont
  {Bloch}}, \bibinfo {author} {\bibfnamefont {J.}~\bibnamefont {Dalibard}}, \
  and\ \bibinfo {author} {\bibfnamefont {S.}~\bibnamefont {Nascimb{\`e}ne}},\
  }\bibfield  {title} {\enquote {\bibinfo {title} {Quantum simulations with
  ultracold quantum gases},}\ }\href {\doibase 10.1038/nphys2259} {\bibfield
  {journal} {\bibinfo  {journal} {Nature Physics}\ }\textbf {\bibinfo {volume}
  {8}},\ \bibinfo {pages} {267--276} (\bibinfo {year} {2012})}\BibitemShut
  {NoStop}%
\bibitem [{\citenamefont {Aspuru-Guzik}\ and\ \citenamefont
  {Walther}(2012)}]{aspuru-guzik12}%
  \BibitemOpen
  \bibfield  {author} {\bibinfo {author} {\bibfnamefont {A.}~\bibnamefont
  {Aspuru-Guzik}}\ and\ \bibinfo {author} {\bibfnamefont {P.}~\bibnamefont
  {Walther}},\ }\bibfield  {title} {\enquote {\bibinfo {title} {Photonic
  quantum simulators},}\ }\href {\doibase 10.1038/nphys2253} {\bibfield
  {journal} {\bibinfo  {journal} {Nature Physics}\ }\textbf {\bibinfo {volume}
  {8}},\ \bibinfo {pages} {285--291} (\bibinfo {year} {2012})}\BibitemShut
  {NoStop}%
\bibitem [{\citenamefont {Peruzzo}\ \emph {et~al.}(2014)\citenamefont
  {Peruzzo}, \citenamefont {McClean}, \citenamefont {Shadbolt}, \citenamefont
  {Yung}, \citenamefont {Zhou}, \citenamefont {Love}, \citenamefont
  {Aspuru-Guzik},\ and\ \citenamefont {O'Brien}}]{peruzzo14}%
  \BibitemOpen
  \bibfield  {author} {\bibinfo {author} {\bibfnamefont {A.}~\bibnamefont
  {Peruzzo}}, \bibinfo {author} {\bibfnamefont {J.}~\bibnamefont {McClean}},
  \bibinfo {author} {\bibfnamefont {P.}~\bibnamefont {Shadbolt}}, \bibinfo
  {author} {\bibfnamefont {M.-H.}\ \bibnamefont {Yung}}, \bibinfo {author}
  {\bibfnamefont {X.-Q.}\ \bibnamefont {Zhou}}, \bibinfo {author}
  {\bibfnamefont {P.~J.}\ \bibnamefont {Love}}, \bibinfo {author}
  {\bibfnamefont {A.}~\bibnamefont {Aspuru-Guzik}}, \ and\ \bibinfo {author}
  {\bibfnamefont {J.~L.}\ \bibnamefont {O'Brien}},\ }\bibfield  {title}
  {\enquote {\bibinfo {title} {A variational eigenvalue solver on a photonic
  quantum processor},}\ }\href {\doibase 10.1038/ncomms5213} {\bibfield
  {journal} {\bibinfo  {journal} {Nature Communications}\ }\textbf {\bibinfo
  {volume} {5}},\ \bibinfo {pages} {4213} (\bibinfo {year} {2014})}\BibitemShut
  {NoStop}%
\bibitem [{\citenamefont {Hartmann}(2016)}]{hartmann16}%
  \BibitemOpen
  \bibfield  {author} {\bibinfo {author} {\bibfnamefont {M.~J.}\ \bibnamefont
  {Hartmann}},\ }\bibfield  {title} {\enquote {\bibinfo {title} {Quantum
  simulation with interacting photons},}\ }\href {\doibase
  10.1088/2040-8978/18/10/104005} {\bibfield  {journal} {\bibinfo  {journal}
  {Journal of Optics}\ }\textbf {\bibinfo {volume} {18}},\ \bibinfo {pages}
  {104005} (\bibinfo {year} {2016})}\BibitemShut {NoStop}%
\bibitem [{\citenamefont {Noh}\ and\ \citenamefont {Angelakis}(2016)}]{noh16}%
  \BibitemOpen
  \bibfield  {author} {\bibinfo {author} {\bibfnamefont {C.}~\bibnamefont
  {Noh}}\ and\ \bibinfo {author} {\bibfnamefont {D.~G.}\ \bibnamefont
  {Angelakis}},\ }\bibfield  {title} {\enquote {\bibinfo {title} {Quantum
  simulations and many-body physics with light},}\ }\href {\doibase
  10.1088/0034-4885/80/1/016401} {\bibfield  {journal} {\bibinfo  {journal}
  {Reports on Progress in Physics}\ }\textbf {\bibinfo {volume} {80}},\
  \bibinfo {pages} {016401} (\bibinfo {year} {2016})}\BibitemShut {NoStop}%
\bibitem [{\citenamefont {Harris}\ \emph {et~al.}(2017)\citenamefont {Harris},
  \citenamefont {Steinbrecher}, \citenamefont {Prabhu}, \citenamefont {Lahini},
  \citenamefont {Mower}, \citenamefont {Bunandar}, \citenamefont {Chen},
  \citenamefont {Wong}, \citenamefont {Baehr-Jones}, \citenamefont {Hochberg},
  \citenamefont {Lloyd},\ and\ \citenamefont {Englund}}]{c_harris17}%
  \BibitemOpen
  \bibfield  {author} {\bibinfo {author} {\bibfnamefont {N.~C.}\ \bibnamefont
  {Harris}}, \bibinfo {author} {\bibfnamefont {G.~R.}\ \bibnamefont
  {Steinbrecher}}, \bibinfo {author} {\bibfnamefont {M.}~\bibnamefont
  {Prabhu}}, \bibinfo {author} {\bibfnamefont {Y.}~\bibnamefont {Lahini}},
  \bibinfo {author} {\bibfnamefont {J.}~\bibnamefont {Mower}}, \bibinfo
  {author} {\bibfnamefont {D.}~\bibnamefont {Bunandar}}, \bibinfo {author}
  {\bibfnamefont {C.}~\bibnamefont {Chen}}, \bibinfo {author} {\bibfnamefont
  {F.~N.~C.}\ \bibnamefont {Wong}}, \bibinfo {author} {\bibfnamefont
  {T.}~\bibnamefont {Baehr-Jones}}, \bibinfo {author} {\bibfnamefont
  {M.}~\bibnamefont {Hochberg}}, \bibinfo {author} {\bibfnamefont
  {S.}~\bibnamefont {Lloyd}}, \ and\ \bibinfo {author} {\bibfnamefont
  {D.}~\bibnamefont {Englund}},\ }\bibfield  {title} {\enquote {\bibinfo
  {title} {Quantum transport simulations in a programmable nanophotonic
  processor},}\ }\href {\doibase 10.1038/nphoton.2017.95} {\bibfield  {journal}
  {\bibinfo  {journal} {Nature Photonics}\ }\textbf {\bibinfo {volume} {11}},\
  \bibinfo {pages} {447--452} (\bibinfo {year} {2017})}\BibitemShut {NoStop}%
\bibitem [{\citenamefont {Berloff}\ \emph {et~al.}(2017)\citenamefont
  {Berloff}, \citenamefont {Silva}, \citenamefont {Kalinin}, \citenamefont
  {Askitopoulos}, \citenamefont {T{\"o}pfer}, \citenamefont {Cilibrizzi},
  \citenamefont {Langbein},\ and\ \citenamefont {Lagoudakis}}]{berloff17}%
  \BibitemOpen
  \bibfield  {author} {\bibinfo {author} {\bibfnamefont {N.~G.}\ \bibnamefont
  {Berloff}}, \bibinfo {author} {\bibfnamefont {M.}~\bibnamefont {Silva}},
  \bibinfo {author} {\bibfnamefont {K.}~\bibnamefont {Kalinin}}, \bibinfo
  {author} {\bibfnamefont {A.}~\bibnamefont {Askitopoulos}}, \bibinfo {author}
  {\bibfnamefont {J.~D.}\ \bibnamefont {T{\"o}pfer}}, \bibinfo {author}
  {\bibfnamefont {P.}~\bibnamefont {Cilibrizzi}}, \bibinfo {author}
  {\bibfnamefont {W.}~\bibnamefont {Langbein}}, \ and\ \bibinfo {author}
  {\bibfnamefont {P.~G.}\ \bibnamefont {Lagoudakis}},\ }\bibfield  {title}
  {\enquote {\bibinfo {title} {Realizing the classical xy hamiltonian in
  polariton simulators},}\ }\href {\doibase 10.1038/nmat4971} {\bibfield
  {journal} {\bibinfo  {journal} {Nature Materials}\ }\textbf {\bibinfo
  {volume} {16}},\ \bibinfo {pages} {1120--1126} (\bibinfo {year}
  {2017})}\BibitemShut {NoStop}%
\bibitem [{\citenamefont {Cirac}\ and\ \citenamefont {Zoller}(1995)}]{cirac95}%
  \BibitemOpen
  \bibfield  {author} {\bibinfo {author} {\bibfnamefont {J.~I.}\ \bibnamefont
  {Cirac}}\ and\ \bibinfo {author} {\bibfnamefont {P.}~\bibnamefont {Zoller}},\
  }\bibfield  {title} {\enquote {\bibinfo {title} {Quantum computations with
  cold trapped ions},}\ }\href {\doibase 10.1103/PhysRevLett.74.4091}
  {\bibfield  {journal} {\bibinfo  {journal} {Phys. Rev. Lett.}\ }\textbf
  {\bibinfo {volume} {74}},\ \bibinfo {pages} {4091--4094} (\bibinfo {year}
  {1995})}\BibitemShut {NoStop}%
\bibitem [{\citenamefont {James}(1998)}]{james98}%
  \BibitemOpen
  \bibfield  {author} {\bibinfo {author} {\bibfnamefont {D.~F.~V.}\
  \bibnamefont {James}},\ }\bibfield  {title} {\enquote {\bibinfo {title}
  {Quantum dynamics of cold trapped ions with application to quantum
  computation},}\ }\href {\doibase 10.1007/s003400050373} {\bibfield  {journal}
  {\bibinfo  {journal} {Applied Physics B}\ }\textbf {\bibinfo {volume} {66}},\
  \bibinfo {pages} {181--190} (\bibinfo {year} {1998})}\BibitemShut {NoStop}%
\bibitem [{\citenamefont {Porras}\ and\ \citenamefont
  {Cirac}(2004)}]{porras04}%
  \BibitemOpen
  \bibfield  {author} {\bibinfo {author} {\bibfnamefont {D.}~\bibnamefont
  {Porras}}\ and\ \bibinfo {author} {\bibfnamefont {J.~I.}\ \bibnamefont
  {Cirac}},\ }\bibfield  {title} {\enquote {\bibinfo {title} {Effective quantum
  spin systems with trapped ions},}\ }\href {\doibase
  10.1103/PhysRevLett.92.207901} {\bibfield  {journal} {\bibinfo  {journal}
  {Phys. Rev. Lett.}\ }\textbf {\bibinfo {volume} {92}},\ \bibinfo {pages}
  {207901} (\bibinfo {year} {2004})}\BibitemShut {NoStop}%
\bibitem [{\citenamefont {Friedenauer}\ \emph {et~al.}(2008)\citenamefont
  {Friedenauer}, \citenamefont {Schmitz}, \citenamefont {Glueckert},
  \citenamefont {Porras},\ and\ \citenamefont {Schaetz}}]{friedenauer08}%
  \BibitemOpen
  \bibfield  {author} {\bibinfo {author} {\bibfnamefont {A.}~\bibnamefont
  {Friedenauer}}, \bibinfo {author} {\bibfnamefont {H.}~\bibnamefont
  {Schmitz}}, \bibinfo {author} {\bibfnamefont {J.~T.}\ \bibnamefont
  {Glueckert}}, \bibinfo {author} {\bibfnamefont {D.}~\bibnamefont {Porras}}, \
  and\ \bibinfo {author} {\bibfnamefont {T.}~\bibnamefont {Schaetz}},\
  }\bibfield  {title} {\enquote {\bibinfo {title} {Simulating a quantum magnet
  with trapped ions},}\ }\href {\doibase 10.1038/nphys1032} {\bibfield
  {journal} {\bibinfo  {journal} {Nature Physics}\ }\textbf {\bibinfo {volume}
  {4}},\ \bibinfo {pages} {757--761} (\bibinfo {year} {2008})}\BibitemShut
  {NoStop}%
\bibitem [{\citenamefont {Häffner}, \citenamefont {Roos},\ and\ \citenamefont
  {Blatt}(2008)}]{haffner08}%
  \BibitemOpen
  \bibfield  {author} {\bibinfo {author} {\bibfnamefont {H.}~\bibnamefont
  {Häffner}}, \bibinfo {author} {\bibfnamefont {C.}~\bibnamefont {Roos}}, \
  and\ \bibinfo {author} {\bibfnamefont {R.}~\bibnamefont {Blatt}},\ }\bibfield
   {title} {\enquote {\bibinfo {title} {Quantum computing with trapped ions},}\
  }\href {\doibase https://doi.org/10.1016/j.physrep.2008.09.003} {\bibfield
  {journal} {\bibinfo  {journal} {Physics Reports}\ }\textbf {\bibinfo {volume}
  {469}},\ \bibinfo {pages} {155--203} (\bibinfo {year} {2008})}\BibitemShut
  {NoStop}%
\bibitem [{\citenamefont {Kim}\ \emph {et~al.}(2010)\citenamefont {Kim},
  \citenamefont {Chang}, \citenamefont {Korenblit}, \citenamefont {Islam},
  \citenamefont {Edwards}, \citenamefont {Freericks}, \citenamefont {Lin},
  \citenamefont {Duan},\ and\ \citenamefont {Monroe}}]{kim10}%
  \BibitemOpen
  \bibfield  {author} {\bibinfo {author} {\bibfnamefont {K.}~\bibnamefont
  {Kim}}, \bibinfo {author} {\bibfnamefont {M.-S.}\ \bibnamefont {Chang}},
  \bibinfo {author} {\bibfnamefont {S.}~\bibnamefont {Korenblit}}, \bibinfo
  {author} {\bibfnamefont {R.}~\bibnamefont {Islam}}, \bibinfo {author}
  {\bibfnamefont {E.~E.}\ \bibnamefont {Edwards}}, \bibinfo {author}
  {\bibfnamefont {J.~K.}\ \bibnamefont {Freericks}}, \bibinfo {author}
  {\bibfnamefont {G.-D.}\ \bibnamefont {Lin}}, \bibinfo {author} {\bibfnamefont
  {L.-M.}\ \bibnamefont {Duan}}, \ and\ \bibinfo {author} {\bibfnamefont
  {C.}~\bibnamefont {Monroe}},\ }\bibfield  {title} {\enquote {\bibinfo {title}
  {Quantum simulation of frustrated ising spins with trapped ions},}\ }\href
  {\doibase 10.1038/nature09071} {\bibfield  {journal} {\bibinfo  {journal}
  {Nature}\ }\textbf {\bibinfo {volume} {465}},\ \bibinfo {pages} {590--593}
  (\bibinfo {year} {2010})}\BibitemShut {NoStop}%
\bibitem [{\citenamefont {Barreiro}\ \emph {et~al.}(2011)\citenamefont
  {Barreiro}, \citenamefont {M{\"u}ller}, \citenamefont {Schindler},
  \citenamefont {Nigg}, \citenamefont {Monz}, \citenamefont {Chwalla},
  \citenamefont {Hennrich}, \citenamefont {Roos}, \citenamefont {Zoller},\ and\
  \citenamefont {Blatt}}]{barreiro11}%
  \BibitemOpen
  \bibfield  {author} {\bibinfo {author} {\bibfnamefont {J.~T.}\ \bibnamefont
  {Barreiro}}, \bibinfo {author} {\bibfnamefont {M.}~\bibnamefont
  {M{\"u}ller}}, \bibinfo {author} {\bibfnamefont {P.}~\bibnamefont
  {Schindler}}, \bibinfo {author} {\bibfnamefont {D.}~\bibnamefont {Nigg}},
  \bibinfo {author} {\bibfnamefont {T.}~\bibnamefont {Monz}}, \bibinfo {author}
  {\bibfnamefont {M.}~\bibnamefont {Chwalla}}, \bibinfo {author} {\bibfnamefont
  {M.}~\bibnamefont {Hennrich}}, \bibinfo {author} {\bibfnamefont {C.~F.}\
  \bibnamefont {Roos}}, \bibinfo {author} {\bibfnamefont {P.}~\bibnamefont
  {Zoller}}, \ and\ \bibinfo {author} {\bibfnamefont {R.}~\bibnamefont
  {Blatt}},\ }\bibfield  {title} {\enquote {\bibinfo {title} {An open-system
  quantum simulator with trapped ions},}\ }\href {\doibase 10.1038/nature09801}
  {\bibfield  {journal} {\bibinfo  {journal} {Nature}\ }\textbf {\bibinfo
  {volume} {470}},\ \bibinfo {pages} {486--491} (\bibinfo {year}
  {2011})}\BibitemShut {NoStop}%
\bibitem [{\citenamefont {Cai}\ \emph {et~al.}(2013)\citenamefont {Cai},
  \citenamefont {Retzker}, \citenamefont {Jelezko},\ and\ \citenamefont
  {Plenio}}]{j_cai13}%
  \BibitemOpen
  \bibfield  {author} {\bibinfo {author} {\bibfnamefont {J.}~\bibnamefont
  {Cai}}, \bibinfo {author} {\bibfnamefont {A.}~\bibnamefont {Retzker}},
  \bibinfo {author} {\bibfnamefont {F.}~\bibnamefont {Jelezko}}, \ and\
  \bibinfo {author} {\bibfnamefont {M.~B.}\ \bibnamefont {Plenio}},\ }\bibfield
   {title} {\enquote {\bibinfo {title} {A large-scale quantum simulator on a
  diamond surface at room temperature},}\ }\href {\doibase 10.1038/nphys2519}
  {\bibfield  {journal} {\bibinfo  {journal} {Nature Physics}\ }\textbf
  {\bibinfo {volume} {9}},\ \bibinfo {pages} {168--173} (\bibinfo {year}
  {2013})}\BibitemShut {NoStop}%
\bibitem [{\citenamefont {Macha}\ \emph {et~al.}(2014)\citenamefont {Macha},
  \citenamefont {Oelsner}, \citenamefont {Reiner}, \citenamefont {Marthaler},
  \citenamefont {Andr{\'e}}, \citenamefont {Sch{\"o}n}, \citenamefont
  {H{\"u}bner}, \citenamefont {Meyer}, \citenamefont {Il'ichev},\ and\
  \citenamefont {Ustinov}}]{macha14}%
  \BibitemOpen
  \bibfield  {author} {\bibinfo {author} {\bibfnamefont {P.}~\bibnamefont
  {Macha}}, \bibinfo {author} {\bibfnamefont {G.}~\bibnamefont {Oelsner}},
  \bibinfo {author} {\bibfnamefont {J.-M.}\ \bibnamefont {Reiner}}, \bibinfo
  {author} {\bibfnamefont {M.}~\bibnamefont {Marthaler}}, \bibinfo {author}
  {\bibfnamefont {S.}~\bibnamefont {Andr{\'e}}}, \bibinfo {author}
  {\bibfnamefont {G.}~\bibnamefont {Sch{\"o}n}}, \bibinfo {author}
  {\bibfnamefont {U.}~\bibnamefont {H{\"u}bner}}, \bibinfo {author}
  {\bibfnamefont {H.-G.}\ \bibnamefont {Meyer}}, \bibinfo {author}
  {\bibfnamefont {E.}~\bibnamefont {Il'ichev}}, \ and\ \bibinfo {author}
  {\bibfnamefont {A.~V.}\ \bibnamefont {Ustinov}},\ }\bibfield  {title}
  {\enquote {\bibinfo {title} {Implementation of a quantum metamaterial using
  superconducting qubits},}\ }\href {\doibase 10.1038/ncomms6146} {\bibfield
  {journal} {\bibinfo  {journal} {Nature Communications}\ }\textbf {\bibinfo
  {volume} {5}},\ \bibinfo {pages} {5146} (\bibinfo {year} {2014})}\BibitemShut
  {NoStop}%
\bibitem [{\citenamefont {Yan}\ \emph {et~al.}(2019)\citenamefont {Yan},
  \citenamefont {Zhang}, \citenamefont {Gong}, \citenamefont {Wu},
  \citenamefont {Zheng}, \citenamefont {Li}, \citenamefont {Wang},
  \citenamefont {Liang}, \citenamefont {Lin}, \citenamefont {Xu}, \citenamefont
  {Guo}, \citenamefont {Sun}, \citenamefont {Peng}, \citenamefont {Xia},
  \citenamefont {Deng}, \citenamefont {Rong}, \citenamefont {You},
  \citenamefont {Nori}, \citenamefont {Fan}, \citenamefont {Zhu},\ and\
  \citenamefont {Pan}}]{yan19}%
  \BibitemOpen
  \bibfield  {author} {\bibinfo {author} {\bibfnamefont {Z.}~\bibnamefont
  {Yan}}, \bibinfo {author} {\bibfnamefont {Y.-R.}\ \bibnamefont {Zhang}},
  \bibinfo {author} {\bibfnamefont {M.}~\bibnamefont {Gong}}, \bibinfo {author}
  {\bibfnamefont {Y.}~\bibnamefont {Wu}}, \bibinfo {author} {\bibfnamefont
  {Y.}~\bibnamefont {Zheng}}, \bibinfo {author} {\bibfnamefont
  {S.}~\bibnamefont {Li}}, \bibinfo {author} {\bibfnamefont {C.}~\bibnamefont
  {Wang}}, \bibinfo {author} {\bibfnamefont {F.}~\bibnamefont {Liang}},
  \bibinfo {author} {\bibfnamefont {J.}~\bibnamefont {Lin}}, \bibinfo {author}
  {\bibfnamefont {Y.}~\bibnamefont {Xu}}, \bibinfo {author} {\bibfnamefont
  {C.}~\bibnamefont {Guo}}, \bibinfo {author} {\bibfnamefont {L.}~\bibnamefont
  {Sun}}, \bibinfo {author} {\bibfnamefont {C.-Z.}\ \bibnamefont {Peng}},
  \bibinfo {author} {\bibfnamefont {K.}~\bibnamefont {Xia}}, \bibinfo {author}
  {\bibfnamefont {H.}~\bibnamefont {Deng}}, \bibinfo {author} {\bibfnamefont
  {H.}~\bibnamefont {Rong}}, \bibinfo {author} {\bibfnamefont {J.~Q.}\
  \bibnamefont {You}}, \bibinfo {author} {\bibfnamefont {F.}~\bibnamefont
  {Nori}}, \bibinfo {author} {\bibfnamefont {H.}~\bibnamefont {Fan}}, \bibinfo
  {author} {\bibfnamefont {X.}~\bibnamefont {Zhu}}, \ and\ \bibinfo {author}
  {\bibfnamefont {J.-W.}\ \bibnamefont {Pan}},\ }\bibfield  {title} {\enquote
  {\bibinfo {title} {Strongly correlated quantum walks with a 12-qubit
  superconducting processor},}\ }\href {\doibase 10.1126/science.aaw1611}
  {\bibfield  {journal} {\bibinfo  {journal} {Science}\ }\textbf {\bibinfo
  {volume} {364}},\ \bibinfo {pages} {753--756} (\bibinfo {year} {2019})},\
  \Eprint
  {http://arxiv.org/abs/https://science.sciencemag.org/content/364/6442/753.full.pdf}
  {https://science.sciencemag.org/content/364/6442/753.full.pdf} \BibitemShut
  {NoStop}%
\bibitem [{\citenamefont {Tan}\ \emph {et~al.}(2019)\citenamefont {Tan},
  \citenamefont {Zhang}, \citenamefont {Yang}, \citenamefont {Chu},
  \citenamefont {Zhu}, \citenamefont {Li}, \citenamefont {Yang}, \citenamefont
  {Song}, \citenamefont {Han}, \citenamefont {Li}, \citenamefont {Dong},
  \citenamefont {Yu}, \citenamefont {Yan}, \citenamefont {Zhu},\ and\
  \citenamefont {Yu}}]{tan19}%
  \BibitemOpen
  \bibfield  {author} {\bibinfo {author} {\bibfnamefont {X.}~\bibnamefont
  {Tan}}, \bibinfo {author} {\bibfnamefont {D.-W.}\ \bibnamefont {Zhang}},
  \bibinfo {author} {\bibfnamefont {Z.}~\bibnamefont {Yang}}, \bibinfo {author}
  {\bibfnamefont {J.}~\bibnamefont {Chu}}, \bibinfo {author} {\bibfnamefont
  {Y.-Q.}\ \bibnamefont {Zhu}}, \bibinfo {author} {\bibfnamefont
  {D.}~\bibnamefont {Li}}, \bibinfo {author} {\bibfnamefont {X.}~\bibnamefont
  {Yang}}, \bibinfo {author} {\bibfnamefont {S.}~\bibnamefont {Song}}, \bibinfo
  {author} {\bibfnamefont {Z.}~\bibnamefont {Han}}, \bibinfo {author}
  {\bibfnamefont {Z.}~\bibnamefont {Li}}, \bibinfo {author} {\bibfnamefont
  {Y.}~\bibnamefont {Dong}}, \bibinfo {author} {\bibfnamefont {H.-F.}\
  \bibnamefont {Yu}}, \bibinfo {author} {\bibfnamefont {H.}~\bibnamefont
  {Yan}}, \bibinfo {author} {\bibfnamefont {S.-L.}\ \bibnamefont {Zhu}}, \ and\
  \bibinfo {author} {\bibfnamefont {Y.}~\bibnamefont {Yu}},\ }\bibfield
  {title} {\enquote {\bibinfo {title} {Experimental measurement of the quantum
  metric tensor and related topological phase transition with a superconducting
  qubit},}\ }\href {\doibase 10.1103/PhysRevLett.122.210401} {\bibfield
  {journal} {\bibinfo  {journal} {Phys. Rev. Lett.}\ }\textbf {\bibinfo
  {volume} {122}},\ \bibinfo {pages} {210401} (\bibinfo {year}
  {2019})}\BibitemShut {NoStop}%
\bibitem [{\citenamefont {Cai}\ \emph {et~al.}(2019)\citenamefont {Cai},
  \citenamefont {Han}, \citenamefont {Mei}, \citenamefont {Xu}, \citenamefont
  {Ma}, \citenamefont {Li}, \citenamefont {Wang}, \citenamefont {Song},
  \citenamefont {Xue}, \citenamefont {Yin}, \citenamefont {Jia},\ and\
  \citenamefont {Sun}}]{w_cai19}%
  \BibitemOpen
  \bibfield  {author} {\bibinfo {author} {\bibfnamefont {W.}~\bibnamefont
  {Cai}}, \bibinfo {author} {\bibfnamefont {J.}~\bibnamefont {Han}}, \bibinfo
  {author} {\bibfnamefont {F.}~\bibnamefont {Mei}}, \bibinfo {author}
  {\bibfnamefont {Y.}~\bibnamefont {Xu}}, \bibinfo {author} {\bibfnamefont
  {Y.}~\bibnamefont {Ma}}, \bibinfo {author} {\bibfnamefont {X.}~\bibnamefont
  {Li}}, \bibinfo {author} {\bibfnamefont {H.}~\bibnamefont {Wang}}, \bibinfo
  {author} {\bibfnamefont {Y.~P.}\ \bibnamefont {Song}}, \bibinfo {author}
  {\bibfnamefont {Z.-Y.}\ \bibnamefont {Xue}}, \bibinfo {author} {\bibfnamefont
  {Z.-q.}\ \bibnamefont {Yin}}, \bibinfo {author} {\bibfnamefont
  {S.}~\bibnamefont {Jia}}, \ and\ \bibinfo {author} {\bibfnamefont
  {L.}~\bibnamefont {Sun}},\ }\bibfield  {title} {\enquote {\bibinfo {title}
  {Observation of topological magnon insulator states in a superconducting
  circuit},}\ }\href {\doibase 10.1103/PhysRevLett.123.080501} {\bibfield
  {journal} {\bibinfo  {journal} {Phys. Rev. Lett.}\ }\textbf {\bibinfo
  {volume} {123}},\ \bibinfo {pages} {080501} (\bibinfo {year}
  {2019})}\BibitemShut {NoStop}%
\bibitem [{\citenamefont {Xu}\ \emph {et~al.}(2020)\citenamefont {Xu},
  \citenamefont {Sun}, \citenamefont {Liu}, \citenamefont {Zhang},
  \citenamefont {Li}, \citenamefont {Dong}, \citenamefont {Ren}, \citenamefont
  {Zhang}, \citenamefont {Nori}, \citenamefont {Zheng}, \citenamefont {Fan},\
  and\ \citenamefont {Wang}}]{xu20}%
  \BibitemOpen
  \bibfield  {author} {\bibinfo {author} {\bibfnamefont {K.}~\bibnamefont
  {Xu}}, \bibinfo {author} {\bibfnamefont {Z.-H.}\ \bibnamefont {Sun}},
  \bibinfo {author} {\bibfnamefont {W.}~\bibnamefont {Liu}}, \bibinfo {author}
  {\bibfnamefont {Y.-R.}\ \bibnamefont {Zhang}}, \bibinfo {author}
  {\bibfnamefont {H.}~\bibnamefont {Li}}, \bibinfo {author} {\bibfnamefont
  {H.}~\bibnamefont {Dong}}, \bibinfo {author} {\bibfnamefont {W.}~\bibnamefont
  {Ren}}, \bibinfo {author} {\bibfnamefont {P.}~\bibnamefont {Zhang}}, \bibinfo
  {author} {\bibfnamefont {F.}~\bibnamefont {Nori}}, \bibinfo {author}
  {\bibfnamefont {D.}~\bibnamefont {Zheng}}, \bibinfo {author} {\bibfnamefont
  {H.}~\bibnamefont {Fan}}, \ and\ \bibinfo {author} {\bibfnamefont
  {H.}~\bibnamefont {Wang}},\ }\bibfield  {title} {\enquote {\bibinfo {title}
  {Probing dynamical phase transitions with a superconducting quantum
  simulator},}\ }\href {\doibase 10.1126/sciadv.aba4935} {\bibfield  {journal}
  {\bibinfo  {journal} {Science Advances}\ }\textbf {\bibinfo {volume} {6}}
  (\bibinfo {year} {2020}),\ 10.1126/sciadv.aba4935},\ \Eprint
  {http://arxiv.org/abs/https://advances.sciencemag.org/content/6/25/eaba4935.full.pdf}
  {https://advances.sciencemag.org/content/6/25/eaba4935.full.pdf} \BibitemShut
  {NoStop}%
\bibitem [{\ \emph {et~al.}(2020), \citenamefont {Arute}, \citenamefont {Arya},
  \citenamefont {Babbush}, \citenamefont {Bacon}, \citenamefont {Bardin},
  \citenamefont {Barends}, \citenamefont {Boixo}, \citenamefont {Broughton},
  \citenamefont {Buckley}, \citenamefont {Buell}, \citenamefont {Burkett},
  \citenamefont {Bushnell}, \citenamefont {Chen}, \citenamefont {Chen},
  \citenamefont {Chiaro}, \citenamefont {Collins}, \citenamefont {Courtney},
  \citenamefont {Demura}, \citenamefont {Dunsworth}, \citenamefont {Farhi},
  \citenamefont {Fowler}, \citenamefont {Foxen}, \citenamefont {Gidney},
  \citenamefont {Giustina}, \citenamefont {Graff}, \citenamefont {Habegger},
  \citenamefont {Harrigan}, \citenamefont {Ho}, \citenamefont {Hong},
  \citenamefont {Huang}, \citenamefont {Huggins}, \citenamefont {Ioffe},
  \citenamefont {Isakov}, \citenamefont {Jeffrey}, \citenamefont {Jiang},
  \citenamefont {Jones}, \citenamefont {Kafri}, \citenamefont {Kechedzhi},
  \citenamefont {Kelly}, \citenamefont {Kim}, \citenamefont {Klimov},
  \citenamefont {Korotkov}, \citenamefont {Kostritsa}, \citenamefont
  {Landhuis}, \citenamefont {Laptev}, \citenamefont {Lindmark}, \citenamefont
  {Lucero}, \citenamefont {Martin}, \citenamefont {Martinis}, \citenamefont
  {McClean}, \citenamefont {McEwen}, \citenamefont {Megrant}, \citenamefont
  {Mi}, \citenamefont {Mohseni}, \citenamefont {Mruczkiewicz}, \citenamefont
  {Mutus}, \citenamefont {Naaman}, \citenamefont {Neeley}, \citenamefont
  {Neill}, \citenamefont {Neven}, \citenamefont {Niu}, \citenamefont
  {O{\textquoteright}Brien}, \citenamefont {Ostby}, \citenamefont {Petukhov},
  \citenamefont {Putterman}, \citenamefont {Quintana}, \citenamefont {Roushan},
  \citenamefont {Rubin}, \citenamefont {Sank}, \citenamefont {Satzinger},
  \citenamefont {Smelyanskiy}, \citenamefont {Strain}, \citenamefont {Sung},
  \citenamefont {Szalay}, \citenamefont {Takeshita}, \citenamefont
  {Vainsencher}, \citenamefont {White}, \citenamefont {Wiebe}, \citenamefont
  {Yao}, \citenamefont {Yeh},\ and\ \citenamefont {Zalcman}}]{google20}%
  \BibitemOpen
  \bibfield  {author} {, \bibinfo {author} {\bibfnamefont {F.}~\bibnamefont
  {Arute}}, \bibinfo {author} {\bibfnamefont {K.}~\bibnamefont {Arya}},
  \bibinfo {author} {\bibfnamefont {R.}~\bibnamefont {Babbush}}, \bibinfo
  {author} {\bibfnamefont {D.}~\bibnamefont {Bacon}}, \bibinfo {author}
  {\bibfnamefont {J.~C.}\ \bibnamefont {Bardin}}, \bibinfo {author}
  {\bibfnamefont {R.}~\bibnamefont {Barends}}, \bibinfo {author} {\bibfnamefont
  {S.}~\bibnamefont {Boixo}}, \bibinfo {author} {\bibfnamefont
  {M.}~\bibnamefont {Broughton}}, \bibinfo {author} {\bibfnamefont {B.~B.}\
  \bibnamefont {Buckley}}, \bibinfo {author} {\bibfnamefont {D.~A.}\
  \bibnamefont {Buell}}, \bibinfo {author} {\bibfnamefont {B.}~\bibnamefont
  {Burkett}}, \bibinfo {author} {\bibfnamefont {N.}~\bibnamefont {Bushnell}},
  \bibinfo {author} {\bibfnamefont {Y.}~\bibnamefont {Chen}}, \bibinfo {author}
  {\bibfnamefont {Z.}~\bibnamefont {Chen}}, \bibinfo {author} {\bibfnamefont
  {B.}~\bibnamefont {Chiaro}}, \bibinfo {author} {\bibfnamefont
  {R.}~\bibnamefont {Collins}}, \bibinfo {author} {\bibfnamefont
  {W.}~\bibnamefont {Courtney}}, \bibinfo {author} {\bibfnamefont
  {S.}~\bibnamefont {Demura}}, \bibinfo {author} {\bibfnamefont
  {A.}~\bibnamefont {Dunsworth}}, \bibinfo {author} {\bibfnamefont
  {E.}~\bibnamefont {Farhi}}, \bibinfo {author} {\bibfnamefont
  {A.}~\bibnamefont {Fowler}}, \bibinfo {author} {\bibfnamefont
  {B.}~\bibnamefont {Foxen}}, \bibinfo {author} {\bibfnamefont
  {C.}~\bibnamefont {Gidney}}, \bibinfo {author} {\bibfnamefont
  {M.}~\bibnamefont {Giustina}}, \bibinfo {author} {\bibfnamefont
  {R.}~\bibnamefont {Graff}}, \bibinfo {author} {\bibfnamefont
  {S.}~\bibnamefont {Habegger}}, \bibinfo {author} {\bibfnamefont {M.~P.}\
  \bibnamefont {Harrigan}}, \bibinfo {author} {\bibfnamefont {A.}~\bibnamefont
  {Ho}}, \bibinfo {author} {\bibfnamefont {S.}~\bibnamefont {Hong}}, \bibinfo
  {author} {\bibfnamefont {T.}~\bibnamefont {Huang}}, \bibinfo {author}
  {\bibfnamefont {W.~J.}\ \bibnamefont {Huggins}}, \bibinfo {author}
  {\bibfnamefont {L.}~\bibnamefont {Ioffe}}, \bibinfo {author} {\bibfnamefont
  {S.~V.}\ \bibnamefont {Isakov}}, \bibinfo {author} {\bibfnamefont
  {E.}~\bibnamefont {Jeffrey}}, \bibinfo {author} {\bibfnamefont
  {Z.}~\bibnamefont {Jiang}}, \bibinfo {author} {\bibfnamefont
  {C.}~\bibnamefont {Jones}}, \bibinfo {author} {\bibfnamefont
  {D.}~\bibnamefont {Kafri}}, \bibinfo {author} {\bibfnamefont
  {K.}~\bibnamefont {Kechedzhi}}, \bibinfo {author} {\bibfnamefont
  {J.}~\bibnamefont {Kelly}}, \bibinfo {author} {\bibfnamefont
  {S.}~\bibnamefont {Kim}}, \bibinfo {author} {\bibfnamefont {P.~V.}\
  \bibnamefont {Klimov}}, \bibinfo {author} {\bibfnamefont {A.}~\bibnamefont
  {Korotkov}}, \bibinfo {author} {\bibfnamefont {F.}~\bibnamefont {Kostritsa}},
  \bibinfo {author} {\bibfnamefont {D.}~\bibnamefont {Landhuis}}, \bibinfo
  {author} {\bibfnamefont {P.}~\bibnamefont {Laptev}}, \bibinfo {author}
  {\bibfnamefont {M.}~\bibnamefont {Lindmark}}, \bibinfo {author}
  {\bibfnamefont {E.}~\bibnamefont {Lucero}}, \bibinfo {author} {\bibfnamefont
  {O.}~\bibnamefont {Martin}}, \bibinfo {author} {\bibfnamefont {J.~M.}\
  \bibnamefont {Martinis}}, \bibinfo {author} {\bibfnamefont {J.~R.}\
  \bibnamefont {McClean}}, \bibinfo {author} {\bibfnamefont {M.}~\bibnamefont
  {McEwen}}, \bibinfo {author} {\bibfnamefont {A.}~\bibnamefont {Megrant}},
  \bibinfo {author} {\bibfnamefont {X.}~\bibnamefont {Mi}}, \bibinfo {author}
  {\bibfnamefont {M.}~\bibnamefont {Mohseni}}, \bibinfo {author} {\bibfnamefont
  {W.}~\bibnamefont {Mruczkiewicz}}, \bibinfo {author} {\bibfnamefont
  {J.}~\bibnamefont {Mutus}}, \bibinfo {author} {\bibfnamefont
  {O.}~\bibnamefont {Naaman}}, \bibinfo {author} {\bibfnamefont
  {M.}~\bibnamefont {Neeley}}, \bibinfo {author} {\bibfnamefont
  {C.}~\bibnamefont {Neill}}, \bibinfo {author} {\bibfnamefont
  {H.}~\bibnamefont {Neven}}, \bibinfo {author} {\bibfnamefont {M.~Y.}\
  \bibnamefont {Niu}}, \bibinfo {author} {\bibfnamefont {T.~E.}\ \bibnamefont
  {O{\textquoteright}Brien}}, \bibinfo {author} {\bibfnamefont
  {E.}~\bibnamefont {Ostby}}, \bibinfo {author} {\bibfnamefont
  {A.}~\bibnamefont {Petukhov}}, \bibinfo {author} {\bibfnamefont
  {H.}~\bibnamefont {Putterman}}, \bibinfo {author} {\bibfnamefont
  {C.}~\bibnamefont {Quintana}}, \bibinfo {author} {\bibfnamefont
  {P.}~\bibnamefont {Roushan}}, \bibinfo {author} {\bibfnamefont {N.~C.}\
  \bibnamefont {Rubin}}, \bibinfo {author} {\bibfnamefont {D.}~\bibnamefont
  {Sank}}, \bibinfo {author} {\bibfnamefont {K.~J.}\ \bibnamefont {Satzinger}},
  \bibinfo {author} {\bibfnamefont {V.}~\bibnamefont {Smelyanskiy}}, \bibinfo
  {author} {\bibfnamefont {D.}~\bibnamefont {Strain}}, \bibinfo {author}
  {\bibfnamefont {K.~J.}\ \bibnamefont {Sung}}, \bibinfo {author}
  {\bibfnamefont {M.}~\bibnamefont {Szalay}}, \bibinfo {author} {\bibfnamefont
  {T.~Y.}\ \bibnamefont {Takeshita}}, \bibinfo {author} {\bibfnamefont
  {A.}~\bibnamefont {Vainsencher}}, \bibinfo {author} {\bibfnamefont
  {T.}~\bibnamefont {White}}, \bibinfo {author} {\bibfnamefont
  {N.}~\bibnamefont {Wiebe}}, \bibinfo {author} {\bibfnamefont {Z.~J.}\
  \bibnamefont {Yao}}, \bibinfo {author} {\bibfnamefont {P.}~\bibnamefont
  {Yeh}}, \ and\ \bibinfo {author} {\bibfnamefont {A.}~\bibnamefont
  {Zalcman}},\ }\bibfield  {title} {\enquote {\bibinfo {title} {Hartree-fock on
  a superconducting qubit quantum computer},}\ }\href {\doibase
  10.1126/science.abb9811} {\bibfield  {journal} {\bibinfo  {journal}
  {Science}\ }\textbf {\bibinfo {volume} {369}},\ \bibinfo {pages} {1084--1089}
  (\bibinfo {year} {2020})},\ \Eprint
  {http://arxiv.org/abs/https://science.sciencemag.org/content/369/6507/1084.full.pdf}
  {https://science.sciencemag.org/content/369/6507/1084.full.pdf} \BibitemShut
  {NoStop}%
\bibitem [{\citenamefont {Harris}\ \emph
  {et~al.}(2010{\natexlab{a}})\citenamefont {Harris}, \citenamefont {Johnson},
  \citenamefont {Lanting}, \citenamefont {Berkley}, \citenamefont {Johansson},
  \citenamefont {Bunyk}, \citenamefont {Tolkacheva}, \citenamefont
  {Ladizinsky}, \citenamefont {Ladizinsky}, \citenamefont {Oh}, \citenamefont
  {Cioata}, \citenamefont {Perminov}, \citenamefont {Spear}, \citenamefont
  {Enderud}, \citenamefont {Rich}, \citenamefont {Uchaikin}, \citenamefont
  {Thom}, \citenamefont {Chapple}, \citenamefont {Wang}, \citenamefont
  {Wilson}, \citenamefont {Amin}, \citenamefont {Dickson}, \citenamefont
  {Karimi}, \citenamefont {Macready}, \citenamefont {Truncik},\ and\
  \citenamefont {Rose}}]{r_harris10}%
  \BibitemOpen
  \bibfield  {author} {\bibinfo {author} {\bibfnamefont {R.}~\bibnamefont
  {Harris}}, \bibinfo {author} {\bibfnamefont {M.~W.}\ \bibnamefont {Johnson}},
  \bibinfo {author} {\bibfnamefont {T.}~\bibnamefont {Lanting}}, \bibinfo
  {author} {\bibfnamefont {A.~J.}\ \bibnamefont {Berkley}}, \bibinfo {author}
  {\bibfnamefont {J.}~\bibnamefont {Johansson}}, \bibinfo {author}
  {\bibfnamefont {P.}~\bibnamefont {Bunyk}}, \bibinfo {author} {\bibfnamefont
  {E.}~\bibnamefont {Tolkacheva}}, \bibinfo {author} {\bibfnamefont
  {E.}~\bibnamefont {Ladizinsky}}, \bibinfo {author} {\bibfnamefont
  {N.}~\bibnamefont {Ladizinsky}}, \bibinfo {author} {\bibfnamefont
  {T.}~\bibnamefont {Oh}}, \bibinfo {author} {\bibfnamefont {F.}~\bibnamefont
  {Cioata}}, \bibinfo {author} {\bibfnamefont {I.}~\bibnamefont {Perminov}},
  \bibinfo {author} {\bibfnamefont {P.}~\bibnamefont {Spear}}, \bibinfo
  {author} {\bibfnamefont {C.}~\bibnamefont {Enderud}}, \bibinfo {author}
  {\bibfnamefont {C.}~\bibnamefont {Rich}}, \bibinfo {author} {\bibfnamefont
  {S.}~\bibnamefont {Uchaikin}}, \bibinfo {author} {\bibfnamefont {M.~C.}\
  \bibnamefont {Thom}}, \bibinfo {author} {\bibfnamefont {E.~M.}\ \bibnamefont
  {Chapple}}, \bibinfo {author} {\bibfnamefont {J.}~\bibnamefont {Wang}},
  \bibinfo {author} {\bibfnamefont {B.}~\bibnamefont {Wilson}}, \bibinfo
  {author} {\bibfnamefont {M.~H.~S.}\ \bibnamefont {Amin}}, \bibinfo {author}
  {\bibfnamefont {N.}~\bibnamefont {Dickson}}, \bibinfo {author} {\bibfnamefont
  {K.}~\bibnamefont {Karimi}}, \bibinfo {author} {\bibfnamefont
  {B.}~\bibnamefont {Macready}}, \bibinfo {author} {\bibfnamefont {C.~J.~S.}\
  \bibnamefont {Truncik}}, \ and\ \bibinfo {author} {\bibfnamefont
  {G.}~\bibnamefont {Rose}},\ }\bibfield  {title} {\enquote {\bibinfo {title}
  {Experimental investigation of an eight-qubit unit cell in a superconducting
  optimization processor},}\ }\href {\doibase 10.1103/PhysRevB.82.024511}
  {\bibfield  {journal} {\bibinfo  {journal} {Phys. Rev. B}\ }\textbf {\bibinfo
  {volume} {82}},\ \bibinfo {pages} {024511} (\bibinfo {year}
  {2010}{\natexlab{a}})}\BibitemShut {NoStop}%
\bibitem [{\citenamefont {Harris}\ \emph
  {et~al.}(2010{\natexlab{b}})\citenamefont {Harris}, \citenamefont {Johnson},
  \citenamefont {Lanting}, \citenamefont {Berkley}, \citenamefont {Johansson},
  \citenamefont {Bunyk}, \citenamefont {Tolkacheva}, \citenamefont
  {Ladizinsky}, \citenamefont {Ladizinsky}, \citenamefont {Oh}, \citenamefont
  {Cioata}, \citenamefont {Perminov}, \citenamefont {Spear}, \citenamefont
  {Enderud}, \citenamefont {Rich}, \citenamefont {Uchaikin}, \citenamefont
  {Thom}, \citenamefont {Chapple}, \citenamefont {Wang}, \citenamefont
  {Wilson}, \citenamefont {Amin}, \citenamefont {Dickson}, \citenamefont
  {Karimi}, \citenamefont {Macready}, \citenamefont {Truncik},\ and\
  \citenamefont {Rose}}]{r_harris10_2}%
  \BibitemOpen
  \bibfield  {author} {\bibinfo {author} {\bibfnamefont {R.}~\bibnamefont
  {Harris}}, \bibinfo {author} {\bibfnamefont {M.~W.}\ \bibnamefont {Johnson}},
  \bibinfo {author} {\bibfnamefont {T.}~\bibnamefont {Lanting}}, \bibinfo
  {author} {\bibfnamefont {A.~J.}\ \bibnamefont {Berkley}}, \bibinfo {author}
  {\bibfnamefont {J.}~\bibnamefont {Johansson}}, \bibinfo {author}
  {\bibfnamefont {P.}~\bibnamefont {Bunyk}}, \bibinfo {author} {\bibfnamefont
  {E.}~\bibnamefont {Tolkacheva}}, \bibinfo {author} {\bibfnamefont
  {E.}~\bibnamefont {Ladizinsky}}, \bibinfo {author} {\bibfnamefont
  {N.}~\bibnamefont {Ladizinsky}}, \bibinfo {author} {\bibfnamefont
  {T.}~\bibnamefont {Oh}}, \bibinfo {author} {\bibfnamefont {F.}~\bibnamefont
  {Cioata}}, \bibinfo {author} {\bibfnamefont {I.}~\bibnamefont {Perminov}},
  \bibinfo {author} {\bibfnamefont {P.}~\bibnamefont {Spear}}, \bibinfo
  {author} {\bibfnamefont {C.}~\bibnamefont {Enderud}}, \bibinfo {author}
  {\bibfnamefont {C.}~\bibnamefont {Rich}}, \bibinfo {author} {\bibfnamefont
  {S.}~\bibnamefont {Uchaikin}}, \bibinfo {author} {\bibfnamefont {M.~C.}\
  \bibnamefont {Thom}}, \bibinfo {author} {\bibfnamefont {E.~M.}\ \bibnamefont
  {Chapple}}, \bibinfo {author} {\bibfnamefont {J.}~\bibnamefont {Wang}},
  \bibinfo {author} {\bibfnamefont {B.}~\bibnamefont {Wilson}}, \bibinfo
  {author} {\bibfnamefont {M.~H.~S.}\ \bibnamefont {Amin}}, \bibinfo {author}
  {\bibfnamefont {N.}~\bibnamefont {Dickson}}, \bibinfo {author} {\bibfnamefont
  {K.}~\bibnamefont {Karimi}}, \bibinfo {author} {\bibfnamefont
  {B.}~\bibnamefont {Macready}}, \bibinfo {author} {\bibfnamefont {C.~J.~S.}\
  \bibnamefont {Truncik}}, \ and\ \bibinfo {author} {\bibfnamefont
  {G.}~\bibnamefont {Rose}},\ }\bibfield  {title} {\enquote {\bibinfo {title}
  {Experimental investigation of an eight-qubit unit cell in a superconducting
  optimization processor},}\ }\href {\doibase 10.1103/PhysRevB.82.024511}
  {\bibfield  {journal} {\bibinfo  {journal} {Phys. Rev. B}\ }\textbf {\bibinfo
  {volume} {82}},\ \bibinfo {pages} {024511} (\bibinfo {year}
  {2010}{\natexlab{b}})}\BibitemShut {NoStop}%
\bibitem [{\citenamefont {Dickson}\ \emph {et~al.}(2013)\citenamefont
  {Dickson}, \citenamefont {Johnson}, \citenamefont {Amin}, \citenamefont
  {Harris}, \citenamefont {Altomare}, \citenamefont {Berkley}, \citenamefont
  {Bunyk}, \citenamefont {Cai}, \citenamefont {Chapple}, \citenamefont
  {Chavez}, \citenamefont {Cioata}, \citenamefont {Cirip}, \citenamefont
  {deBuen}, \citenamefont {Drew-Brook}, \citenamefont {Enderud}, \citenamefont
  {Gildert}, \citenamefont {Hamze}, \citenamefont {Hilton}, \citenamefont
  {Hoskinson}, \citenamefont {Karimi}, \citenamefont {Ladizinsky},
  \citenamefont {Ladizinsky}, \citenamefont {Lanting}, \citenamefont {Mahon},
  \citenamefont {Neufeld}, \citenamefont {Oh}, \citenamefont {Perminov},
  \citenamefont {Petroff}, \citenamefont {Przybysz}, \citenamefont {Rich},
  \citenamefont {Spear}, \citenamefont {Tcaciuc}, \citenamefont {Thom},
  \citenamefont {Tolkacheva}, \citenamefont {Uchaikin}, \citenamefont {Wang},
  \citenamefont {Wilson}, \citenamefont {Merali},\ and\ \citenamefont
  {Rose}}]{dickson13}%
  \BibitemOpen
  \bibfield  {author} {\bibinfo {author} {\bibfnamefont {N.~G.}\ \bibnamefont
  {Dickson}}, \bibinfo {author} {\bibfnamefont {M.~W.}\ \bibnamefont
  {Johnson}}, \bibinfo {author} {\bibfnamefont {M.~H.}\ \bibnamefont {Amin}},
  \bibinfo {author} {\bibfnamefont {R.}~\bibnamefont {Harris}}, \bibinfo
  {author} {\bibfnamefont {F.}~\bibnamefont {Altomare}}, \bibinfo {author}
  {\bibfnamefont {A.~J.}\ \bibnamefont {Berkley}}, \bibinfo {author}
  {\bibfnamefont {P.}~\bibnamefont {Bunyk}}, \bibinfo {author} {\bibfnamefont
  {J.}~\bibnamefont {Cai}}, \bibinfo {author} {\bibfnamefont {E.~M.}\
  \bibnamefont {Chapple}}, \bibinfo {author} {\bibfnamefont {P.}~\bibnamefont
  {Chavez}}, \bibinfo {author} {\bibfnamefont {F.}~\bibnamefont {Cioata}},
  \bibinfo {author} {\bibfnamefont {T.}~\bibnamefont {Cirip}}, \bibinfo
  {author} {\bibfnamefont {P.}~\bibnamefont {deBuen}}, \bibinfo {author}
  {\bibfnamefont {M.}~\bibnamefont {Drew-Brook}}, \bibinfo {author}
  {\bibfnamefont {C.}~\bibnamefont {Enderud}}, \bibinfo {author} {\bibfnamefont
  {S.}~\bibnamefont {Gildert}}, \bibinfo {author} {\bibfnamefont
  {F.}~\bibnamefont {Hamze}}, \bibinfo {author} {\bibfnamefont {J.~P.}\
  \bibnamefont {Hilton}}, \bibinfo {author} {\bibfnamefont {E.}~\bibnamefont
  {Hoskinson}}, \bibinfo {author} {\bibfnamefont {K.}~\bibnamefont {Karimi}},
  \bibinfo {author} {\bibfnamefont {E.}~\bibnamefont {Ladizinsky}}, \bibinfo
  {author} {\bibfnamefont {N.}~\bibnamefont {Ladizinsky}}, \bibinfo {author}
  {\bibfnamefont {T.}~\bibnamefont {Lanting}}, \bibinfo {author} {\bibfnamefont
  {T.}~\bibnamefont {Mahon}}, \bibinfo {author} {\bibfnamefont
  {R.}~\bibnamefont {Neufeld}}, \bibinfo {author} {\bibfnamefont
  {T.}~\bibnamefont {Oh}}, \bibinfo {author} {\bibfnamefont {I.}~\bibnamefont
  {Perminov}}, \bibinfo {author} {\bibfnamefont {C.}~\bibnamefont {Petroff}},
  \bibinfo {author} {\bibfnamefont {A.}~\bibnamefont {Przybysz}}, \bibinfo
  {author} {\bibfnamefont {C.}~\bibnamefont {Rich}}, \bibinfo {author}
  {\bibfnamefont {P.}~\bibnamefont {Spear}}, \bibinfo {author} {\bibfnamefont
  {A.}~\bibnamefont {Tcaciuc}}, \bibinfo {author} {\bibfnamefont {M.~C.}\
  \bibnamefont {Thom}}, \bibinfo {author} {\bibfnamefont {E.}~\bibnamefont
  {Tolkacheva}}, \bibinfo {author} {\bibfnamefont {S.}~\bibnamefont
  {Uchaikin}}, \bibinfo {author} {\bibfnamefont {J.}~\bibnamefont {Wang}},
  \bibinfo {author} {\bibfnamefont {A.~B.}\ \bibnamefont {Wilson}}, \bibinfo
  {author} {\bibfnamefont {Z.}~\bibnamefont {Merali}}, \ and\ \bibinfo {author}
  {\bibfnamefont {G.}~\bibnamefont {Rose}},\ }\bibfield  {title} {\enquote
  {\bibinfo {title} {Thermally assisted quantum annealing of a 16-qubit
  problem},}\ }\href {\doibase 10.1038/ncomms2920} {\bibfield  {journal}
  {\bibinfo  {journal} {Nature Communications}\ }\textbf {\bibinfo {volume}
  {4}},\ \bibinfo {pages} {1903} (\bibinfo {year} {2013})}\BibitemShut
  {NoStop}%
\bibitem [{\citenamefont {Johnson}\ \emph {et~al.}(2011)\citenamefont
  {Johnson}, \citenamefont {Amin}, \citenamefont {Gildert}, \citenamefont
  {Lanting}, \citenamefont {Hamze}, \citenamefont {Dickson}, \citenamefont
  {Harris}, \citenamefont {Berkley}, \citenamefont {Johansson}, \citenamefont
  {Bunyk}, \citenamefont {Chapple}, \citenamefont {Enderud}, \citenamefont
  {Hilton}, \citenamefont {Karimi}, \citenamefont {Ladizinsky}, \citenamefont
  {Ladizinsky}, \citenamefont {Oh}, \citenamefont {Perminov}, \citenamefont
  {Rich}, \citenamefont {Thom}, \citenamefont {Tolkacheva}, \citenamefont
  {Truncik}, \citenamefont {Uchaikin}, \citenamefont {Wang}, \citenamefont
  {Wilson},\ and\ \citenamefont {Rose}}]{johnson11}%
  \BibitemOpen
  \bibfield  {author} {\bibinfo {author} {\bibfnamefont {M.~W.}\ \bibnamefont
  {Johnson}}, \bibinfo {author} {\bibfnamefont {M.~H.~S.}\ \bibnamefont
  {Amin}}, \bibinfo {author} {\bibfnamefont {S.}~\bibnamefont {Gildert}},
  \bibinfo {author} {\bibfnamefont {T.}~\bibnamefont {Lanting}}, \bibinfo
  {author} {\bibfnamefont {F.}~\bibnamefont {Hamze}}, \bibinfo {author}
  {\bibfnamefont {N.}~\bibnamefont {Dickson}}, \bibinfo {author} {\bibfnamefont
  {R.}~\bibnamefont {Harris}}, \bibinfo {author} {\bibfnamefont {A.~J.}\
  \bibnamefont {Berkley}}, \bibinfo {author} {\bibfnamefont {J.}~\bibnamefont
  {Johansson}}, \bibinfo {author} {\bibfnamefont {P.}~\bibnamefont {Bunyk}},
  \bibinfo {author} {\bibfnamefont {E.~M.}\ \bibnamefont {Chapple}}, \bibinfo
  {author} {\bibfnamefont {C.}~\bibnamefont {Enderud}}, \bibinfo {author}
  {\bibfnamefont {J.~P.}\ \bibnamefont {Hilton}}, \bibinfo {author}
  {\bibfnamefont {K.}~\bibnamefont {Karimi}}, \bibinfo {author} {\bibfnamefont
  {E.}~\bibnamefont {Ladizinsky}}, \bibinfo {author} {\bibfnamefont
  {N.}~\bibnamefont {Ladizinsky}}, \bibinfo {author} {\bibfnamefont
  {T.}~\bibnamefont {Oh}}, \bibinfo {author} {\bibfnamefont {I.}~\bibnamefont
  {Perminov}}, \bibinfo {author} {\bibfnamefont {C.}~\bibnamefont {Rich}},
  \bibinfo {author} {\bibfnamefont {M.~C.}\ \bibnamefont {Thom}}, \bibinfo
  {author} {\bibfnamefont {E.}~\bibnamefont {Tolkacheva}}, \bibinfo {author}
  {\bibfnamefont {C.~J.~S.}\ \bibnamefont {Truncik}}, \bibinfo {author}
  {\bibfnamefont {S.}~\bibnamefont {Uchaikin}}, \bibinfo {author}
  {\bibfnamefont {J.}~\bibnamefont {Wang}}, \bibinfo {author} {\bibfnamefont
  {B.}~\bibnamefont {Wilson}}, \ and\ \bibinfo {author} {\bibfnamefont
  {G.}~\bibnamefont {Rose}},\ }\bibfield  {title} {\enquote {\bibinfo {title}
  {Quantum annealing with manufactured spins},}\ }\href {\doibase
  10.1038/nature10012} {\bibfield  {journal} {\bibinfo  {journal} {Nature}\
  }\textbf {\bibinfo {volume} {473}},\ \bibinfo {pages} {194--198} (\bibinfo
  {year} {2011})}\BibitemShut {NoStop}%
\bibitem [{\citenamefont {Harris}\ \emph {et~al.}(2018)\citenamefont {Harris},
  \citenamefont {Sato}, \citenamefont {Berkley}, \citenamefont {Reis},
  \citenamefont {Altomare}, \citenamefont {Amin}, \citenamefont {Boothby},
  \citenamefont {Bunyk}, \citenamefont {Deng}, \citenamefont {Enderud},
  \citenamefont {Huang}, \citenamefont {Hoskinson}, \citenamefont {Johnson},
  \citenamefont {Ladizinsky}, \citenamefont {Ladizinsky}, \citenamefont
  {Lanting}, \citenamefont {Li}, \citenamefont {Medina}, \citenamefont
  {Molavi}, \citenamefont {Neufeld}, \citenamefont {Oh}, \citenamefont
  {Pavlov}, \citenamefont {Perminov}, \citenamefont {Poulin-Lamarre},
  \citenamefont {Rich}, \citenamefont {Smirnov}, \citenamefont {Swenson},
  \citenamefont {Tsai}, \citenamefont {Volkmann}, \citenamefont {Whittaker},\
  and\ \citenamefont {Yao}}]{harris18}%
  \BibitemOpen
  \bibfield  {author} {\bibinfo {author} {\bibfnamefont {R.}~\bibnamefont
  {Harris}}, \bibinfo {author} {\bibfnamefont {Y.}~\bibnamefont {Sato}},
  \bibinfo {author} {\bibfnamefont {A.~J.}\ \bibnamefont {Berkley}}, \bibinfo
  {author} {\bibfnamefont {M.}~\bibnamefont {Reis}}, \bibinfo {author}
  {\bibfnamefont {F.}~\bibnamefont {Altomare}}, \bibinfo {author}
  {\bibfnamefont {M.~H.}\ \bibnamefont {Amin}}, \bibinfo {author}
  {\bibfnamefont {K.}~\bibnamefont {Boothby}}, \bibinfo {author} {\bibfnamefont
  {P.}~\bibnamefont {Bunyk}}, \bibinfo {author} {\bibfnamefont
  {C.}~\bibnamefont {Deng}}, \bibinfo {author} {\bibfnamefont {C.}~\bibnamefont
  {Enderud}}, \bibinfo {author} {\bibfnamefont {S.}~\bibnamefont {Huang}},
  \bibinfo {author} {\bibfnamefont {E.}~\bibnamefont {Hoskinson}}, \bibinfo
  {author} {\bibfnamefont {M.~W.}\ \bibnamefont {Johnson}}, \bibinfo {author}
  {\bibfnamefont {E.}~\bibnamefont {Ladizinsky}}, \bibinfo {author}
  {\bibfnamefont {N.}~\bibnamefont {Ladizinsky}}, \bibinfo {author}
  {\bibfnamefont {T.}~\bibnamefont {Lanting}}, \bibinfo {author} {\bibfnamefont
  {R.}~\bibnamefont {Li}}, \bibinfo {author} {\bibfnamefont {T.}~\bibnamefont
  {Medina}}, \bibinfo {author} {\bibfnamefont {R.}~\bibnamefont {Molavi}},
  \bibinfo {author} {\bibfnamefont {R.}~\bibnamefont {Neufeld}}, \bibinfo
  {author} {\bibfnamefont {T.}~\bibnamefont {Oh}}, \bibinfo {author}
  {\bibfnamefont {I.}~\bibnamefont {Pavlov}}, \bibinfo {author} {\bibfnamefont
  {I.}~\bibnamefont {Perminov}}, \bibinfo {author} {\bibfnamefont
  {G.}~\bibnamefont {Poulin-Lamarre}}, \bibinfo {author} {\bibfnamefont
  {C.}~\bibnamefont {Rich}}, \bibinfo {author} {\bibfnamefont {A.}~\bibnamefont
  {Smirnov}}, \bibinfo {author} {\bibfnamefont {L.}~\bibnamefont {Swenson}},
  \bibinfo {author} {\bibfnamefont {N.}~\bibnamefont {Tsai}}, \bibinfo {author}
  {\bibfnamefont {M.}~\bibnamefont {Volkmann}}, \bibinfo {author}
  {\bibfnamefont {J.}~\bibnamefont {Whittaker}}, \ and\ \bibinfo {author}
  {\bibfnamefont {J.}~\bibnamefont {Yao}},\ }\bibfield  {title} {\enquote
  {\bibinfo {title} {Phase transitions in a programmable quantum spin glass
  simulator},}\ }\href {\doibase 10.1126/science.aat2025} {\bibfield  {journal}
  {\bibinfo  {journal} {Science}\ }\textbf {\bibinfo {volume} {361}},\ \bibinfo
  {pages} {162--165} (\bibinfo {year} {2018})},\ \Eprint
  {http://arxiv.org/abs/https://science.sciencemag.org/content/361/6398/162.full.pdf}
  {https://science.sciencemag.org/content/361/6398/162.full.pdf} \BibitemShut
  {NoStop}%
\bibitem [{\citenamefont {Gardas}\ \emph {et~al.}(2018)\citenamefont {Gardas},
  \citenamefont {Dziarmaga}, \citenamefont {Zurek},\ and\ \citenamefont
  {Zwolak}}]{gardas18}%
  \BibitemOpen
  \bibfield  {author} {\bibinfo {author} {\bibfnamefont {B.}~\bibnamefont
  {Gardas}}, \bibinfo {author} {\bibfnamefont {J.}~\bibnamefont {Dziarmaga}},
  \bibinfo {author} {\bibfnamefont {W.~H.}\ \bibnamefont {Zurek}}, \ and\
  \bibinfo {author} {\bibfnamefont {M.}~\bibnamefont {Zwolak}},\ }\bibfield
  {title} {\enquote {\bibinfo {title} {Defects in quantum computers},}\ }\href
  {\doibase 10.1038/s41598-018-22763-2} {\bibfield  {journal} {\bibinfo
  {journal} {Scientific Reports}\ }\textbf {\bibinfo {volume} {8}},\ \bibinfo
  {pages} {4539} (\bibinfo {year} {2018})}\BibitemShut {NoStop}%
\bibitem [{\citenamefont {Weinberg}\ \emph {et~al.}(2020)\citenamefont
  {Weinberg}, \citenamefont {Tylutki}, \citenamefont {R\"onkk\"o},
  \citenamefont {Westerholm}, \citenamefont {\AA{}str\"om}, \citenamefont
  {Manninen}, \citenamefont {T\"orm\"a},\ and\ \citenamefont
  {Sandvik}}]{weinberg20}%
  \BibitemOpen
  \bibfield  {author} {\bibinfo {author} {\bibfnamefont {P.}~\bibnamefont
  {Weinberg}}, \bibinfo {author} {\bibfnamefont {M.}~\bibnamefont {Tylutki}},
  \bibinfo {author} {\bibfnamefont {J.~M.}\ \bibnamefont {R\"onkk\"o}},
  \bibinfo {author} {\bibfnamefont {J.}~\bibnamefont {Westerholm}}, \bibinfo
  {author} {\bibfnamefont {J.~A.}\ \bibnamefont {\AA{}str\"om}}, \bibinfo
  {author} {\bibfnamefont {P.}~\bibnamefont {Manninen}}, \bibinfo {author}
  {\bibfnamefont {P.}~\bibnamefont {T\"orm\"a}}, \ and\ \bibinfo {author}
  {\bibfnamefont {A.~W.}\ \bibnamefont {Sandvik}},\ }\bibfield  {title}
  {\enquote {\bibinfo {title} {Scaling and diabatic effects in quantum
  annealing with a d-wave device},}\ }\href {\doibase
  10.1103/PhysRevLett.124.090502} {\bibfield  {journal} {\bibinfo  {journal}
  {Phys. Rev. Lett.}\ }\textbf {\bibinfo {volume} {124}},\ \bibinfo {pages}
  {090502} (\bibinfo {year} {2020})}\BibitemShut {NoStop}%
\bibitem [{\citenamefont {Bando}\ \emph {et~al.}(2020)\citenamefont {Bando},
  \citenamefont {Susa}, \citenamefont {Oshiyama}, \citenamefont {Shibata},
  \citenamefont {Ohzeki}, \citenamefont {G\'omez-Ruiz}, \citenamefont {Lidar},
  \citenamefont {Suzuki}, \citenamefont {del Campo},\ and\ \citenamefont
  {Nishimori}}]{bando20}%
  \BibitemOpen
  \bibfield  {author} {\bibinfo {author} {\bibfnamefont {Y.}~\bibnamefont
  {Bando}}, \bibinfo {author} {\bibfnamefont {Y.}~\bibnamefont {Susa}},
  \bibinfo {author} {\bibfnamefont {H.}~\bibnamefont {Oshiyama}}, \bibinfo
  {author} {\bibfnamefont {N.}~\bibnamefont {Shibata}}, \bibinfo {author}
  {\bibfnamefont {M.}~\bibnamefont {Ohzeki}}, \bibinfo {author} {\bibfnamefont
  {F.~J.}\ \bibnamefont {G\'omez-Ruiz}}, \bibinfo {author} {\bibfnamefont
  {D.~A.}\ \bibnamefont {Lidar}}, \bibinfo {author} {\bibfnamefont
  {S.}~\bibnamefont {Suzuki}}, \bibinfo {author} {\bibfnamefont
  {A.}~\bibnamefont {del Campo}}, \ and\ \bibinfo {author} {\bibfnamefont
  {H.}~\bibnamefont {Nishimori}},\ }\bibfield  {title} {\enquote {\bibinfo
  {title} {Probing the universality of topological defect formation in a
  quantum annealer: Kibble-zurek mechanism and beyond},}\ }\href {\doibase
  10.1103/PhysRevResearch.2.033369} {\bibfield  {journal} {\bibinfo  {journal}
  {Phys. Rev. Research}\ }\textbf {\bibinfo {volume} {2}},\ \bibinfo {pages}
  {033369} (\bibinfo {year} {2020})}\BibitemShut {NoStop}%
\bibitem [{\citenamefont {Novoselov}\ \emph {et~al.}(2004)\citenamefont
  {Novoselov}, \citenamefont {Geim}, \citenamefont {Morozov}, \citenamefont
  {Jiang}, \citenamefont {Zhang}, \citenamefont {Dubonos}, \citenamefont
  {Grigorieva},\ and\ \citenamefont
  {Firsov}}]{Novoselov.Electric_field_atomically_thin_carbon}%
  \BibitemOpen
  \bibfield  {author} {\bibinfo {author} {\bibfnamefont {K.~S.}\ \bibnamefont
  {Novoselov}}, \bibinfo {author} {\bibfnamefont {A.~K.}\ \bibnamefont {Geim}},
  \bibinfo {author} {\bibfnamefont {S.~V.}\ \bibnamefont {Morozov}}, \bibinfo
  {author} {\bibfnamefont {D.}~\bibnamefont {Jiang}}, \bibinfo {author}
  {\bibfnamefont {Y.}~\bibnamefont {Zhang}}, \bibinfo {author} {\bibfnamefont
  {S.~V.}\ \bibnamefont {Dubonos}}, \bibinfo {author} {\bibfnamefont {I.~V.}\
  \bibnamefont {Grigorieva}}, \ and\ \bibinfo {author} {\bibfnamefont {A.~A.}\
  \bibnamefont {Firsov}},\ }\bibfield  {title} {\enquote {\bibinfo {title}
  {Electric field effect in atomically thin carbon films},}\ }\href@noop {}
  {\bibfield  {journal} {\bibinfo  {journal} {Science}\ }\textbf {\bibinfo
  {volume} {306}},\ \bibinfo {pages} {666--669} (\bibinfo {year}
  {2004})}\BibitemShut {NoStop}%
\bibitem [{\citenamefont {Novoselov}\ \emph {et~al.}(2005)\citenamefont
  {Novoselov}, \citenamefont {Geim}, \citenamefont {Morozov}, \citenamefont
  {Jiang}, \citenamefont {Katsnelson}, \citenamefont {Grigorieva},
  \citenamefont {Dubonos},\ and\ \citenamefont {Firsov}}]{graphene1b}%
  \BibitemOpen
  \bibfield  {author} {\bibinfo {author} {\bibfnamefont {K.~S.}\ \bibnamefont
  {Novoselov}}, \bibinfo {author} {\bibfnamefont {A.~K.}\ \bibnamefont {Geim}},
  \bibinfo {author} {\bibfnamefont {S.~V.}\ \bibnamefont {Morozov}}, \bibinfo
  {author} {\bibfnamefont {D.}~\bibnamefont {Jiang}}, \bibinfo {author}
  {\bibfnamefont {M.~I.}\ \bibnamefont {Katsnelson}}, \bibinfo {author}
  {\bibfnamefont {I.~V.}\ \bibnamefont {Grigorieva}}, \bibinfo {author}
  {\bibfnamefont {S.~V.}\ \bibnamefont {Dubonos}}, \ and\ \bibinfo {author}
  {\bibfnamefont {A.~A.}\ \bibnamefont {Firsov}},\ }\bibfield  {title}
  {\enquote {\bibinfo {title} {Two-dimensional gas of massless dirac fermions
  in graphene},}\ }\href@noop {} {\bibfield  {journal} {\bibinfo  {journal}
  {Nature}\ }\textbf {\bibinfo {volume} {438}},\ \bibinfo {pages} {197}
  (\bibinfo {year} {2005})}\BibitemShut {NoStop}%
\bibitem [{\citenamefont {Geim}\ and\ \citenamefont
  {Novoselov}(2007)}]{geim2007}%
  \BibitemOpen
  \bibfield  {author} {\bibinfo {author} {\bibfnamefont {A.~K.}\ \bibnamefont
  {Geim}}\ and\ \bibinfo {author} {\bibfnamefont {K.~S.}\ \bibnamefont
  {Novoselov}},\ }\bibfield  {title} {\enquote {\bibinfo {title} {The rise of
  graphene},}\ }\href@noop {} {\bibfield  {journal} {\bibinfo  {journal}
  {Nature. Mater.}\ }\textbf {\bibinfo {volume} {6}},\ \bibinfo {pages} {183}
  (\bibinfo {year} {2007})}\BibitemShut {NoStop}%
\bibitem [{\citenamefont {Nalwa}(2009)}]{Nalwa.MagneticNanostructures}%
  \BibitemOpen
  \bibinfo {editor} {\bibfnamefont {H.~S.}\ \bibnamefont {Nalwa}},\ ed.,\
  \href@noop {} {\emph {\bibinfo {title} {Magnetic Nanostructures}}}\ (\bibinfo
   {publisher} {American Scientific Publishers},\ \bibinfo {year}
  {2009})\BibitemShut {NoStop}%
\bibitem [{\citenamefont {Avouris}, \citenamefont {Chen},\ and\ \citenamefont
  {Perebeinos}(2007)}]{avouris2007}%
  \BibitemOpen
  \bibfield  {author} {\bibinfo {author} {\bibfnamefont {P.}~\bibnamefont
  {Avouris}}, \bibinfo {author} {\bibfnamefont {Z.}~\bibnamefont {Chen}}, \
  and\ \bibinfo {author} {\bibfnamefont {V.}~\bibnamefont {Perebeinos}},\
  }\bibfield  {title} {\enquote {\bibinfo {title} {Carbon-based electronics},}\
  }\href@noop {} {\bibfield  {journal} {\bibinfo  {journal} {Nature Nanotech.}\
  }\textbf {\bibinfo {volume} {2}},\ \bibinfo {pages} {605} (\bibinfo {year}
  {2007})}\BibitemShut {NoStop}%
\bibitem [{\citenamefont {Castro~Neto}\ \emph {et~al.}(2009)\citenamefont
  {Castro~Neto}, \citenamefont {Kotov}, \citenamefont {Nilsson}, \citenamefont
  {Pereira}, \citenamefont {Peres},\ and\ \citenamefont
  {Uchoa}}]{castro2009adatoms}%
  \BibitemOpen
  \bibfield  {author} {\bibinfo {author} {\bibfnamefont {A.~H.}\ \bibnamefont
  {Castro~Neto}}, \bibinfo {author} {\bibfnamefont {V.~N.}\ \bibnamefont
  {Kotov}}, \bibinfo {author} {\bibfnamefont {J.}~\bibnamefont {Nilsson}},
  \bibinfo {author} {\bibfnamefont {V.~M.}\ \bibnamefont {Pereira}}, \bibinfo
  {author} {\bibfnamefont {N.~M.~R.}\ \bibnamefont {Peres}}, \ and\ \bibinfo
  {author} {\bibfnamefont {B.}~\bibnamefont {Uchoa}},\ }\bibfield  {title}
  {\enquote {\bibinfo {title} {Adatoms in graphene},}\ }\href@noop {}
  {\bibfield  {journal} {\bibinfo  {journal} {Solid State Communications}\
  }\textbf {\bibinfo {volume} {149}},\ \bibinfo {pages} {1094--1100} (\bibinfo
  {year} {2009})}\BibitemShut {NoStop}%
\bibitem [{\citenamefont {Fritz}\ and\ \citenamefont {Vojta}(2013)}]{Lars}%
  \BibitemOpen
  \bibfield  {author} {\bibinfo {author} {\bibfnamefont {L.}~\bibnamefont
  {Fritz}}\ and\ \bibinfo {author} {\bibfnamefont {M.}~\bibnamefont {Vojta}},\
  }\href@noop {} {\bibfield  {journal} {\bibinfo  {journal} {Rep. Prog. Phys.}\
  }\textbf {\bibinfo {volume} {76}},\ \bibinfo {pages} {032501} (\bibinfo
  {year} {2013})}\BibitemShut {NoStop}%
\bibitem [{\citenamefont {Shytov}, \citenamefont {Abanin},\ and\ \citenamefont
  {Levitov}(2009)}]{shytov2009long}%
  \BibitemOpen
  \bibfield  {author} {\bibinfo {author} {\bibfnamefont {A.~V.}\ \bibnamefont
  {Shytov}}, \bibinfo {author} {\bibfnamefont {D.~A.}\ \bibnamefont {Abanin}},
  \ and\ \bibinfo {author} {\bibfnamefont {L.~S.}\ \bibnamefont {Levitov}},\
  }\bibfield  {title} {\enquote {\bibinfo {title} {Long-range interaction
  between adatoms in graphene},}\ }\href@noop {} {\bibfield  {journal}
  {\bibinfo  {journal} {Phys. Rev. Lett.}\ }\textbf {\bibinfo {volume} {103}},\
  \bibinfo {pages} {016806} (\bibinfo {year} {2009})}\BibitemShut {NoStop}%
\bibitem [{\citenamefont {Kotov}\ \emph {et~al.}(2012)\citenamefont {Kotov},
  \citenamefont {Uchoa}, \citenamefont {Pereira}, \citenamefont {Guinea},\ and\
  \citenamefont {Castro~Neto}}]{Kotov2012}%
  \BibitemOpen
  \bibfield  {author} {\bibinfo {author} {\bibfnamefont {V.~N.}\ \bibnamefont
  {Kotov}}, \bibinfo {author} {\bibfnamefont {B.}~\bibnamefont {Uchoa}},
  \bibinfo {author} {\bibfnamefont {V.~M.}\ \bibnamefont {Pereira}}, \bibinfo
  {author} {\bibfnamefont {F.}~\bibnamefont {Guinea}}, \ and\ \bibinfo {author}
  {\bibfnamefont {A.~H.}\ \bibnamefont {Castro~Neto}},\ }\bibfield  {title}
  {\enquote {\bibinfo {title} {Electron-electron interactions in graphene:
  Current status and perspectives},}\ }\href@noop {} {\bibfield  {journal}
  {\bibinfo  {journal} {Rev. Mod. Phys.}\ }\textbf {\bibinfo {volume} {84}},\
  \bibinfo {pages} {1067--1125} (\bibinfo {year} {2012})}\BibitemShut {NoStop}%
\bibitem [{\citenamefont {Castro-Neto}\ \emph {et~al.}(2009)\citenamefont
  {Castro-Neto}, \citenamefont {Guinea}, \citenamefont {Peres}, \citenamefont
  {Novoselov},\ and\ \citenamefont {Geim.}}]{graphene_review1}%
  \BibitemOpen
  \bibfield  {author} {\bibinfo {author} {\bibfnamefont {A.~H.}\ \bibnamefont
  {Castro-Neto}}, \bibinfo {author} {\bibfnamefont {F.}~\bibnamefont {Guinea}},
  \bibinfo {author} {\bibfnamefont {N.~M.~R.}\ \bibnamefont {Peres}}, \bibinfo
  {author} {\bibfnamefont {K.~S.}\ \bibnamefont {Novoselov}}, \ and\ \bibinfo
  {author} {\bibfnamefont {A.~K.}\ \bibnamefont {Geim.}},\ }\bibfield  {title}
  {\enquote {\bibinfo {title} {The electronic properties of graphene},}\
  }\href@noop {} {\bibfield  {journal} {\bibinfo  {journal} {Rev. Mod. Phys.}\
  }\textbf {\bibinfo {volume} {81}},\ \bibinfo {pages} {109} (\bibinfo {year}
  {2009})}\BibitemShut {NoStop}%
\bibitem [{\citenamefont {Sarma}\ \emph {et~al.}(2011)\citenamefont {Sarma},
  \citenamefont {Adam}, \citenamefont {Hwang}, ,\ and\ \citenamefont
  {Rossi}}]{graphene_review2}%
  \BibitemOpen
  \bibfield  {author} {\bibinfo {author} {\bibfnamefont {S.~D.}\ \bibnamefont
  {Sarma}}, \bibinfo {author} {\bibfnamefont {S.}~\bibnamefont {Adam}},
  \bibinfo {author} {\bibfnamefont {E.~H.}\ \bibnamefont {Hwang}}, , \ and\
  \bibinfo {author} {\bibfnamefont {E.}~\bibnamefont {Rossi}},\ }\bibfield
  {title} {\enquote {\bibinfo {title} {Electronic transport in two-dimensional
  graphene},}\ }\href@noop {} {\bibfield  {journal} {\bibinfo  {journal} {Rev.
  Mod. Phys.}\ }\textbf {\bibinfo {volume} {83}},\ \bibinfo {pages} {407}
  (\bibinfo {year} {2011})}\BibitemShut {NoStop}%
\bibitem [{\citenamefont {Chen}\ \emph {et~al.}(2011)\citenamefont {Chen},
  \citenamefont {Li}, \citenamefont {Cullen}, \citenamefont {Williams},\ and\
  \citenamefont {Fuhrer}}]{graphene_Kondo}%
  \BibitemOpen
  \bibfield  {author} {\bibinfo {author} {\bibfnamefont {J.-H.}\ \bibnamefont
  {Chen}}, \bibinfo {author} {\bibfnamefont {L.}~\bibnamefont {Li}}, \bibinfo
  {author} {\bibfnamefont {W.~G.}\ \bibnamefont {Cullen}}, \bibinfo {author}
  {\bibfnamefont {E.~D.}\ \bibnamefont {Williams}}, \ and\ \bibinfo {author}
  {\bibfnamefont {M.~S.}\ \bibnamefont {Fuhrer}},\ }\bibfield  {title}
  {\enquote {\bibinfo {title} {Tunable kondo effect in graphene with
  defects.}}\ }\href@noop {} {\bibfield  {journal} {\bibinfo  {journal} {Nature
  Physics 7}\ }\textbf {\bibinfo {volume} {7}},\ \bibinfo {pages} {535}
  (\bibinfo {year} {2011})}\BibitemShut {NoStop}%
\bibitem [{\citenamefont {Cornaglia}, \citenamefont {Usaj},\ and\ \citenamefont
  {Balseiro}(2009)}]{cornaglia2009}%
  \BibitemOpen
  \bibfield  {author} {\bibinfo {author} {\bibfnamefont {P.~S.}\ \bibnamefont
  {Cornaglia}}, \bibinfo {author} {\bibfnamefont {G.}~\bibnamefont {Usaj}}, \
  and\ \bibinfo {author} {\bibfnamefont {C.~A.}\ \bibnamefont {Balseiro}},\
  }\bibfield  {title} {\enquote {\bibinfo {title} {Localized spins on
  graphene},}\ }\href {\doibase 10.1103/PhysRevLett.102.046801} {\bibfield
  {journal} {\bibinfo  {journal} {Phys. Rev. Lett.}\ }\textbf {\bibinfo
  {volume} {102}},\ \bibinfo {pages} {046801} (\bibinfo {year}
  {2009})}\BibitemShut {NoStop}%
\bibitem [{\citenamefont {Sengupta}\ and\ \citenamefont
  {Baskaran}(2008)}]{sengupta2008}%
  \BibitemOpen
  \bibfield  {author} {\bibinfo {author} {\bibfnamefont {K.}~\bibnamefont
  {Sengupta}}\ and\ \bibinfo {author} {\bibfnamefont {G.}~\bibnamefont
  {Baskaran}},\ }\bibfield  {title} {\enquote {\bibinfo {title} {Tuning kondo
  physics in graphene with gate voltage},}\ }\href@noop {} {\bibfield
  {journal} {\bibinfo  {journal} {Phys. Rev. B}\ }\textbf {\bibinfo {volume}
  {77}},\ \bibinfo {pages} {045417} (\bibinfo {year} {2008})}\BibitemShut
  {NoStop}%
\bibitem [{\citenamefont {Jacob}\ and\ \citenamefont
  {Kotliar}(2010)}]{Jacob10}%
  \BibitemOpen
  \bibfield  {author} {\bibinfo {author} {\bibfnamefont {D.}~\bibnamefont
  {Jacob}}\ and\ \bibinfo {author} {\bibfnamefont {G.}~\bibnamefont
  {Kotliar}},\ }\href@noop {} {\bibfield  {journal} {\bibinfo  {journal} {Phys.
  Rev. B}\ }\textbf {\bibinfo {volume} {82}},\ \bibinfo {pages} {085423}
  (\bibinfo {year} {2010})}\BibitemShut {NoStop}%
\bibitem [{\citenamefont {Uchoa}, \citenamefont {Rappoport},\ and\
  \citenamefont {Castro~Neto}(2011)}]{Uchoa2011}%
  \BibitemOpen
  \bibfield  {author} {\bibinfo {author} {\bibfnamefont {B.}~\bibnamefont
  {Uchoa}}, \bibinfo {author} {\bibfnamefont {T.~G.}\ \bibnamefont
  {Rappoport}}, \ and\ \bibinfo {author} {\bibfnamefont {A.~H.}\ \bibnamefont
  {Castro~Neto}},\ }\bibfield  {title} {\enquote {\bibinfo {title} {Kondo
  quantum criticality of magnetic adatoms in graphene},}\ }\href {\doibase
  10.1103/PhysRevLett.106.016801} {\bibfield  {journal} {\bibinfo  {journal}
  {Phys. Rev. Lett.}\ }\textbf {\bibinfo {volume} {106}},\ \bibinfo {pages}
  {016801} (\bibinfo {year} {2011})}\BibitemShut {NoStop}%
\bibitem [{\citenamefont {Thimm}, \citenamefont {Kroha},\ and\ \citenamefont
  {von Delft}(1999)}]{Thimm1999}%
  \BibitemOpen
  \bibfield  {author} {\bibinfo {author} {\bibfnamefont {W.~B.}\ \bibnamefont
  {Thimm}}, \bibinfo {author} {\bibfnamefont {J.}~\bibnamefont {Kroha}}, \ and\
  \bibinfo {author} {\bibfnamefont {J.}~\bibnamefont {von Delft}},\ }\bibfield
  {title} {\enquote {\bibinfo {title} {Kondo box: A magnetic impurity in an
  ultrasmall metallic grain},}\ }\href@noop {} {\bibfield  {journal} {\bibinfo
  {journal} {Phys. Rev. Lett.}\ }\textbf {\bibinfo {volume} {82}},\ \bibinfo
  {pages} {2143--2146} (\bibinfo {year} {1999})}\BibitemShut {NoStop}%
\bibitem [{\citenamefont {Schlottmann}(2001)}]{schlottmann2001kondo}%
  \BibitemOpen
  \bibfield  {author} {\bibinfo {author} {\bibfnamefont {P.}~\bibnamefont
  {Schlottmann}},\ }\bibfield  {title} {\enquote {\bibinfo {title} {Kondo
  effect in a nanosized particle},}\ }\href@noop {} {\bibfield  {journal}
  {\bibinfo  {journal} {Phys. Rev. B}\ }\textbf {\bibinfo {volume} {65}},\
  \bibinfo {pages} {024420} (\bibinfo {year} {2001})}\BibitemShut {NoStop}%
\bibitem [{\citenamefont {Simon}\ and\ \citenamefont
  {Affleck}(2002)}]{simon2002finite}%
  \BibitemOpen
  \bibfield  {author} {\bibinfo {author} {\bibfnamefont {P.}~\bibnamefont
  {Simon}}\ and\ \bibinfo {author} {\bibfnamefont {I.}~\bibnamefont
  {Affleck}},\ }\bibfield  {title} {\enquote {\bibinfo {title} {Finite-size
  effects in conductance measurements on quantum dots},}\ }\href@noop {}
  {\bibfield  {journal} {\bibinfo  {journal} {Phys. Rev. Lett.}\ }\textbf
  {\bibinfo {volume} {89}},\ \bibinfo {pages} {206602} (\bibinfo {year}
  {2002})}\BibitemShut {NoStop}%
\bibitem [{\citenamefont {Simon}\ and\ \citenamefont
  {Affleck}(2003)}]{simon2003kondo}%
  \BibitemOpen
  \bibfield  {author} {\bibinfo {author} {\bibfnamefont {P.}~\bibnamefont
  {Simon}}\ and\ \bibinfo {author} {\bibfnamefont {I.}~\bibnamefont
  {Affleck}},\ }\bibfield  {title} {\enquote {\bibinfo {title} {Kondo screening
  cloud effects in mesoscopic devices},}\ }\href@noop {} {\bibfield  {journal}
  {\bibinfo  {journal} {Phys. Rev. B}\ }\textbf {\bibinfo {volume} {68}},\
  \bibinfo {pages} {115304} (\bibinfo {year} {2003})}\BibitemShut {NoStop}%
\bibitem [{\citenamefont {Hand}, \citenamefont {Kroha},\ and\ \citenamefont
  {Monien}(2006)}]{hand2006spin}%
  \BibitemOpen
  \bibfield  {author} {\bibinfo {author} {\bibfnamefont {T.}~\bibnamefont
  {Hand}}, \bibinfo {author} {\bibfnamefont {J.}~\bibnamefont {Kroha}}, \ and\
  \bibinfo {author} {\bibfnamefont {H.}~\bibnamefont {Monien}},\ }\bibfield
  {title} {\enquote {\bibinfo {title} {Spin correlations and finite-size
  effects in the one-dimensional kondo box},}\ }\href@noop {} {\bibfield
  {journal} {\bibinfo  {journal} {Phys. Rev. Lett.}\ }\textbf {\bibinfo
  {volume} {97}},\ \bibinfo {pages} {136604} (\bibinfo {year}
  {2006})}\BibitemShut {NoStop}%
\bibitem [{\citenamefont {Hanl}\ and\ \citenamefont
  {Weichselbaum}(2014)}]{Hanl2014}%
  \BibitemOpen
  \bibfield  {author} {\bibinfo {author} {\bibfnamefont {M.}~\bibnamefont
  {Hanl}}\ and\ \bibinfo {author} {\bibfnamefont {A.}~\bibnamefont
  {Weichselbaum}},\ }\bibfield  {title} {\enquote {\bibinfo {title} {Local
  susceptibility and kondo scaling in the presence of finite bandwidth},}\
  }\href {\doibase 10.1103/PhysRevB.89.075130} {\bibfield  {journal} {\bibinfo
  {journal} {Phys. Rev. B}\ }\textbf {\bibinfo {volume} {89}},\ \bibinfo
  {pages} {075130} (\bibinfo {year} {2014})}\BibitemShut {NoStop}%
\bibitem [{\citenamefont {Yang}\ and\ \citenamefont
  {Feiguin}(2017)}]{Yang2017}%
  \BibitemOpen
  \bibfield  {author} {\bibinfo {author} {\bibfnamefont {C.}~\bibnamefont
  {Yang}}\ and\ \bibinfo {author} {\bibfnamefont {A.~E.}\ \bibnamefont
  {Feiguin}},\ }\bibfield  {title} {\enquote {\bibinfo {title} {Unveiling the
  internal entanglement structure of the kondo singlet},}\ }\href {\doibase
  10.1103/PhysRevB.95.115106} {\bibfield  {journal} {\bibinfo  {journal} {Phys.
  Rev. B}\ }\textbf {\bibinfo {volume} {95}},\ \bibinfo {pages} {115106}
  (\bibinfo {year} {2017})}\BibitemShut {NoStop}%
\bibitem [{\citenamefont {Debertolis}, \citenamefont {Florens},\ and\
  \citenamefont {Snyman}(2021)}]{debertolis2021fewbody}%
  \BibitemOpen
  \bibfield  {author} {\bibinfo {author} {\bibfnamefont {M.}~\bibnamefont
  {Debertolis}}, \bibinfo {author} {\bibfnamefont {S.}~\bibnamefont {Florens}},
  \ and\ \bibinfo {author} {\bibfnamefont {I.}~\bibnamefont {Snyman}},\
  }\href@noop {} {\enquote {\bibinfo {title} {Few-body nature of a kondo
  correlated ground state},}\ } (\bibinfo {year} {2021}),\ \Eprint
  {http://arxiv.org/abs/2011.12678} {arXiv:2011.12678 [cond-mat.str-el]}
  \BibitemShut {NoStop}%
\bibitem [{\citenamefont {Nakada}\ \emph {et~al.}(1996)\citenamefont {Nakada},
  \citenamefont {Fujita}, \citenamefont {Dresselhaus},\ and\ \citenamefont
  {Dresselhaus}}]{Fujita1996}%
  \BibitemOpen
  \bibfield  {author} {\bibinfo {author} {\bibfnamefont {K.}~\bibnamefont
  {Nakada}}, \bibinfo {author} {\bibfnamefont {M.}~\bibnamefont {Fujita}},
  \bibinfo {author} {\bibfnamefont {G.}~\bibnamefont {Dresselhaus}}, \ and\
  \bibinfo {author} {\bibfnamefont {M.~S.}\ \bibnamefont {Dresselhaus}},\
  }\bibfield  {title} {\enquote {\bibinfo {title} {Edge state in graphene
  ribbons: Nanometer size effect and edge shape dependence},}\ }\href@noop {}
  {\bibfield  {journal} {\bibinfo  {journal} {Phys. Rev. B}\ }\textbf {\bibinfo
  {volume} {54}},\ \bibinfo {pages} {17954--17961} (\bibinfo {year}
  {1996})}\BibitemShut {NoStop}%
\bibitem [{\citenamefont {Kobayashi}\ \emph {et~al.}(2006)\citenamefont
  {Kobayashi}, \citenamefont {Fukui}, \citenamefont {Enoki},\ and\
  \citenamefont {Kusakabe}}]{Kobayashi2006}%
  \BibitemOpen
  \bibfield  {author} {\bibinfo {author} {\bibfnamefont {Y.}~\bibnamefont
  {Kobayashi}}, \bibinfo {author} {\bibfnamefont {K.-i.}\ \bibnamefont
  {Fukui}}, \bibinfo {author} {\bibfnamefont {T.}~\bibnamefont {Enoki}}, \ and\
  \bibinfo {author} {\bibfnamefont {K.}~\bibnamefont {Kusakabe}},\ }\bibfield
  {title} {\enquote {\bibinfo {title} {Edge state on hydrogen-terminated
  graphite edges investigated by scanning tunneling microscopy},}\ }\href@noop
  {} {\bibfield  {journal} {\bibinfo  {journal} {Phys. Rev. B}\ }\textbf
  {\bibinfo {volume} {73}},\ \bibinfo {pages} {125415} (\bibinfo {year}
  {2006})}\BibitemShut {NoStop}%
\bibitem [{\citenamefont {Allerdt}, \citenamefont {Feiguin},\ and\
  \citenamefont {Das~Sarma}(2017)}]{Allerdt2017}%
  \BibitemOpen
  \bibfield  {author} {\bibinfo {author} {\bibfnamefont {A.}~\bibnamefont
  {Allerdt}}, \bibinfo {author} {\bibfnamefont {A.~E.}\ \bibnamefont
  {Feiguin}}, \ and\ \bibinfo {author} {\bibfnamefont {S.}~\bibnamefont
  {Das~Sarma}},\ }\bibfield  {title} {\enquote {\bibinfo {title} {Competition
  between kondo effect and rkky physics in graphene magnetism},}\ }\href
  {\doibase 10.1103/PhysRevB.95.104402} {\bibfield  {journal} {\bibinfo
  {journal} {Phys. Rev. B}\ }\textbf {\bibinfo {volume} {95}},\ \bibinfo
  {pages} {104402} (\bibinfo {year} {2017})}\BibitemShut {NoStop}%
\bibitem [{\citenamefont {Yosida}(1957)}]{Yoisida.Magnetic_properties}%
  \BibitemOpen
  \bibfield  {author} {\bibinfo {author} {\bibfnamefont {K.}~\bibnamefont
  {Yosida}},\ }\bibfield  {title} {\enquote {\bibinfo {title} {Magnetic
  properties of cu-mn alloys},}\ }\href {\doibase 10.1103/PhysRev.106.893}
  {\bibfield  {journal} {\bibinfo  {journal} {Phys. Rev.}\ }\textbf {\bibinfo
  {volume} {106}},\ \bibinfo {pages} {893--898} (\bibinfo {year}
  {1957})}\BibitemShut {NoStop}%
\bibitem [{\citenamefont {Ruderman}\ and\ \citenamefont
  {Kittel}(1954)}]{Kittel.Indirect_exchange}%
  \BibitemOpen
  \bibfield  {author} {\bibinfo {author} {\bibfnamefont {M.~A.}\ \bibnamefont
  {Ruderman}}\ and\ \bibinfo {author} {\bibfnamefont {C.}~\bibnamefont
  {Kittel}},\ }\bibfield  {title} {\enquote {\bibinfo {title} {Indirect
  exchange coupling of nuclear magnetic moments by conduction electrons},}\
  }\href {\doibase 10.1103/PhysRev.96.99} {\bibfield  {journal} {\bibinfo
  {journal} {Phys. Rev.}\ }\textbf {\bibinfo {volume} {96}},\ \bibinfo {pages}
  {99--102} (\bibinfo {year} {1954})}\BibitemShut {NoStop}%
\bibitem [{\citenamefont {Kasuya}(1956)}]{Kasuya.Theory_of_metallic}%
  \BibitemOpen
  \bibfield  {author} {\bibinfo {author} {\bibfnamefont {T.}~\bibnamefont
  {Kasuya}},\ }\bibfield  {title} {\enquote {\bibinfo {title} {A theory of
  metallic ferro- and antiferromagnetism on zener's model},}\ }\href@noop {}
  {\bibfield  {journal} {\bibinfo  {journal} {Progress of Theoretical Physics}\
  }\textbf {\bibinfo {volume} {16}},\ \bibinfo {pages} {45--57} (\bibinfo
  {year} {1956})}\BibitemShut {NoStop}%
\bibitem [{\citenamefont {B\"usser}, \citenamefont {Martins},\ and\
  \citenamefont {Feiguin}(2013)}]{Busser2013}%
  \BibitemOpen
  \bibfield  {author} {\bibinfo {author} {\bibfnamefont {C.~A.}\ \bibnamefont
  {B\"usser}}, \bibinfo {author} {\bibfnamefont {G.~B.}\ \bibnamefont
  {Martins}}, \ and\ \bibinfo {author} {\bibfnamefont {A.~E.}\ \bibnamefont
  {Feiguin}},\ }\bibfield  {title} {\enquote {\bibinfo {title} {Lanczos
  transformation for quantum impurity problems in $d$-dimensional lattices:
  Application to graphene nanoribbons},}\ }\href {\doibase
  10.1103/PhysRevB.88.245113} {\bibfield  {journal} {\bibinfo  {journal} {Phys.
  Rev. B}\ }\textbf {\bibinfo {volume} {88}},\ \bibinfo {pages} {245113}
  (\bibinfo {year} {2013})}\BibitemShut {NoStop}%
\bibitem [{\citenamefont {Allerdt}\ \emph {et~al.}(2015)\citenamefont
  {Allerdt}, \citenamefont {B\"usser}, \citenamefont {Martins},\ and\
  \citenamefont {Feiguin}}]{Allerdt.Kondo}%
  \BibitemOpen
  \bibfield  {author} {\bibinfo {author} {\bibfnamefont {A.}~\bibnamefont
  {Allerdt}}, \bibinfo {author} {\bibfnamefont {C.~A.}\ \bibnamefont
  {B\"usser}}, \bibinfo {author} {\bibfnamefont {G.~B.}\ \bibnamefont
  {Martins}}, \ and\ \bibinfo {author} {\bibfnamefont {A.~E.}\ \bibnamefont
  {Feiguin}},\ }\bibfield  {title} {\enquote {\bibinfo {title} {Kondo versus
  indirect exchange: Role of lattice and actual range of rkky interactions in
  real materials},}\ }\href {\doibase 10.1103/PhysRevB.91.085101} {\bibfield
  {journal} {\bibinfo  {journal} {Phys. Rev. B}\ }\textbf {\bibinfo {volume}
  {91}},\ \bibinfo {pages} {085101} (\bibinfo {year} {2015})}\BibitemShut
  {NoStop}%
\bibitem [{\citenamefont {Allerdt}\ and\ \citenamefont
  {Feiguin}(2019)}]{Frontiers_review}%
  \BibitemOpen
  \bibfield  {author} {\bibinfo {author} {\bibfnamefont {A.}~\bibnamefont
  {Allerdt}}\ and\ \bibinfo {author} {\bibfnamefont {A.~E.}\ \bibnamefont
  {Feiguin}},\ }\bibfield  {title} {\enquote {\bibinfo {title} {A numerically
  exact approach to quantum impurity problems in realistic lattice
  geometries},}\ }\href {\doibase 10.3389/fphy.2019.00067} {\bibfield
  {journal} {\bibinfo  {journal} {Frontiers in Physics}\ }\textbf {\bibinfo
  {volume} {7}},\ \bibinfo {pages} {67} (\bibinfo {year} {2019})}\BibitemShut
  {NoStop}%
\bibitem [{\citenamefont {White}(1992)}]{White1992}%
  \BibitemOpen
  \bibfield  {author} {\bibinfo {author} {\bibfnamefont {S.~R.}\ \bibnamefont
  {White}},\ }\bibfield  {title} {\enquote {\bibinfo {title} {Density matrix
  formulation for quantum renormalization groups.}}\ }\href@noop {} {\bibfield
  {journal} {\bibinfo  {journal} {Phys. Rev. Lett.}\ }\textbf {\bibinfo
  {volume} {69}},\ \bibinfo {pages} {2863--2866} (\bibinfo {year}
  {1992})}\BibitemShut {NoStop}%
\bibitem [{\citenamefont {White}(1993)}]{White1993}%
  \BibitemOpen
  \bibfield  {author} {\bibinfo {author} {\bibfnamefont {S.~R.}\ \bibnamefont
  {White}},\ }\bibfield  {title} {\enquote {\bibinfo {title} {Density-matrix
  algorithms for quantum renormalization groups.}}\ }\href@noop {} {\bibfield
  {journal} {\bibinfo  {journal} {Phys. Rev. B}\ }\textbf {\bibinfo {volume}
  {48}},\ \bibinfo {pages} {10345--10356} (\bibinfo {year} {1993})}\BibitemShut
  {NoStop}%
\bibitem [{\citenamefont {Schollw\"ock}(2005)}]{Schollwock2005}%
  \BibitemOpen
  \bibfield  {author} {\bibinfo {author} {\bibfnamefont {U.}~\bibnamefont
  {Schollw\"ock}},\ }\bibfield  {title} {\enquote {\bibinfo {title} {The
  density-matrix renormalization group},}\ }\href {\doibase
  10.1103/RevModPhys.77.259} {\bibfield  {journal} {\bibinfo  {journal} {Rev.
  Mod. Phys.}\ }\textbf {\bibinfo {volume} {77}},\ \bibinfo {pages} {259--315}
  (\bibinfo {year} {2005})}\BibitemShut {NoStop}%
\bibitem [{\citenamefont {Schollw\"ock}(2011)}]{Schollwock2011}%
  \BibitemOpen
  \bibfield  {author} {\bibinfo {author} {\bibfnamefont {U.}~\bibnamefont
  {Schollw\"ock}},\ }\bibfield  {title} {\enquote {\bibinfo {title} {The
  density-matrix renormalization group in the age of matrix product states},}\
  }\href {\doibase https://doi.org/10.1016/j.aop.2010.09.012} {\bibfield
  {journal} {\bibinfo  {journal} {Annals of Physics}\ }\textbf {\bibinfo
  {volume} {326}},\ \bibinfo {pages} {96 -- 192} (\bibinfo {year} {2011})},\
  \bibinfo {note} {january 2011 Special Issue}\BibitemShut {NoStop}%
\bibitem [{\citenamefont {Feiguin}(2013)}]{Feiguin2013a}%
  \BibitemOpen
  \bibfield  {author} {\bibinfo {author} {\bibfnamefont {A.~E.}\ \bibnamefont
  {Feiguin}},\ }\bibfield  {title} {\enquote {\bibinfo {title} {The density
  matrix renormalization group},}\ }in\ \href@noop {} {\emph {\bibinfo
  {booktitle} {Strongly correlated systems: Numerical methods}}},\ \bibinfo
  {editor} {edited by\ \bibinfo {editor} {\bibfnamefont {A.}~\bibnamefont
  {Avella}}\ and\ \bibinfo {editor} {\bibfnamefont {F.}~\bibnamefont
  {Mancini}}}\ (\bibinfo  {publisher} {Springer, Heidelberg, Berlin},\ \bibinfo
  {year} {2013})\BibitemShut {NoStop}%
\bibitem [{\citenamefont {Loss}\ and\ \citenamefont
  {DiVincenzo}(1998)}]{divincenzo98}%
  \BibitemOpen
  \bibfield  {author} {\bibinfo {author} {\bibfnamefont {D.}~\bibnamefont
  {Loss}}\ and\ \bibinfo {author} {\bibfnamefont {D.~P.}\ \bibnamefont
  {DiVincenzo}},\ }\bibfield  {title} {\enquote {\bibinfo {title} {Quantum
  computation with quantum dots},}\ }\href {\doibase 10.1103/PhysRevA.57.120}
  {\bibfield  {journal} {\bibinfo  {journal} {Phys. Rev. A}\ }\textbf {\bibinfo
  {volume} {57}},\ \bibinfo {pages} {120--126} (\bibinfo {year}
  {1998})}\BibitemShut {NoStop}%
\bibitem [{\citenamefont {Zhang}\ \emph {et~al.}(2009)\citenamefont {Zhang},
  \citenamefont {Shaikh}, \citenamefont {Tsui},\ and\ \citenamefont
  {Swager}}]{Zhang2009}%
  \BibitemOpen
  \bibfield  {author} {\bibinfo {author} {\bibfnamefont {W.}~\bibnamefont
  {Zhang}}, \bibinfo {author} {\bibfnamefont {A.~U.}\ \bibnamefont {Shaikh}},
  \bibinfo {author} {\bibfnamefont {E.~Y.}\ \bibnamefont {Tsui}}, \ and\
  \bibinfo {author} {\bibfnamefont {T.~M.}\ \bibnamefont {Swager}},\ }\bibfield
   {title} {\enquote {\bibinfo {title} {Cobalt porphyrin functionalized carbon
  nanotubes for oxygen reduction},}\ }\href {\doibase 10.1021/cm900747t}
  {\bibfield  {journal} {\bibinfo  {journal} {Chem. Mater.}\ }\textbf {\bibinfo
  {volume} {21}},\ \bibinfo {pages} {3234--3241} (\bibinfo {year}
  {2009})}\BibitemShut {NoStop}%
\bibitem [{\citenamefont {Lee}\ \emph {et~al.}(2011)\citenamefont {Lee},
  \citenamefont {Lee}, \citenamefont {Lee}, \citenamefont {Kim},\ and\
  \citenamefont {Kim}}]{Lee2011}%
  \BibitemOpen
  \bibfield  {author} {\bibinfo {author} {\bibfnamefont {D.~H.}\ \bibnamefont
  {Lee}}, \bibinfo {author} {\bibfnamefont {W.~J.}\ \bibnamefont {Lee}},
  \bibinfo {author} {\bibfnamefont {W.~J.}\ \bibnamefont {Lee}}, \bibinfo
  {author} {\bibfnamefont {S.~O.}\ \bibnamefont {Kim}}, \ and\ \bibinfo
  {author} {\bibfnamefont {Y.-H.}\ \bibnamefont {Kim}},\ }\bibfield  {title}
  {\enquote {\bibinfo {title} {{Theory, Synthesis, and Oxygen Reduction
  Catalysis of Fe-Porphyrin-Like Carbon Nanotube}},}\ }\href {\doibase
  10.1103/PhysRevLett.106.175502} {\bibfield  {journal} {\bibinfo  {journal}
  {Phys. Rev. Lett.}\ }\textbf {\bibinfo {volume} {106}},\ \bibinfo {pages}
  {175502} (\bibinfo {year} {2011})}\BibitemShut {NoStop}%
\bibitem [{\citenamefont {Chung}, \citenamefont {Won},\ and\ \citenamefont
  {Zelenay}(2013)}]{Chung2013}%
  \BibitemOpen
  \bibfield  {author} {\bibinfo {author} {\bibfnamefont {H.~T.}\ \bibnamefont
  {Chung}}, \bibinfo {author} {\bibfnamefont {J.~H.}\ \bibnamefont {Won}}, \
  and\ \bibinfo {author} {\bibfnamefont {P.}~\bibnamefont {Zelenay}},\
  }\bibfield  {title} {\enquote {\bibinfo {title} {{Active and stable carbon
  nanotube/nanoparticle composite electrocatalyst for oxygen reduction.}}}\
  }\href {\doibase 10.1038/ncomms2944} {\bibfield  {journal} {\bibinfo
  {journal} {Nat. Commun.}\ }\textbf {\bibinfo {volume} {4}},\ \bibinfo {pages}
  {1922} (\bibinfo {year} {2013})}\BibitemShut {NoStop}%
\bibitem [{\citenamefont {Zhu}\ and\ \citenamefont {Dong}(2013)}]{Zhu2013}%
  \BibitemOpen
  \bibfield  {author} {\bibinfo {author} {\bibfnamefont {C.}~\bibnamefont
  {Zhu}}\ and\ \bibinfo {author} {\bibfnamefont {S.}~\bibnamefont {Dong}},\
  }\bibfield  {title} {\enquote {\bibinfo {title} {Recent progress in
  graphene-based nanomaterials as advanced electrocatalysts towards oxygen
  reduction reaction},}\ }\href {\doibase 10.1039/c2nr33839d} {\bibfield
  {journal} {\bibinfo  {journal} {Nanoscale}\ }\textbf {\bibinfo {volume}
  {5}},\ \bibinfo {pages} {1753--67} (\bibinfo {year} {2013})}\BibitemShut
  {NoStop}%
\bibitem [{\citenamefont {Jia}\ \emph {et~al.}(2015)\citenamefont {Jia},
  \citenamefont {Ramaswamy}, \citenamefont {Hafiz}, \citenamefont {Tylus},
  \citenamefont {Strickland},\ and\ \citenamefont {Wu}}]{Jia2015}%
  \BibitemOpen
  \bibfield  {author} {\bibinfo {author} {\bibfnamefont {Q.}~\bibnamefont
  {Jia}}, \bibinfo {author} {\bibfnamefont {N.}~\bibnamefont {Ramaswamy}},
  \bibinfo {author} {\bibfnamefont {H.}~\bibnamefont {Hafiz}}, \bibinfo
  {author} {\bibfnamefont {U.}~\bibnamefont {Tylus}}, \bibinfo {author}
  {\bibfnamefont {K.}~\bibnamefont {Strickland}}, \ and\ \bibinfo {author}
  {\bibfnamefont {G.}~\bibnamefont {Wu}},\ }\bibfield  {title} {\enquote
  {\bibinfo {title} {Experimental observation of redox-induced fe--n switching
  behavior as a determinant role for oxygen reduction activity},}\ }\href
  {\doibase 10.1021/acsnano.5b05984} {\bibfield  {journal} {\bibinfo  {journal}
  {ACS Nano}\ }\textbf {\bibinfo {volume} {9}},\ \bibinfo {pages}
  {12496--12505} (\bibinfo {year} {2015})}\BibitemShut {NoStop}%
\bibitem [{\citenamefont {Jia}\ \emph {et~al.}(2016)\citenamefont {Jia},
  \citenamefont {Ramaswamy}, \citenamefont {Tylus}, \citenamefont {Strickland},
  \citenamefont {Li}, \citenamefont {Serov}, \citenamefont {Artyushkova},
  \citenamefont {Atanassov}, \citenamefont {Anibal}, \citenamefont {Gumeci},
  \citenamefont {Barton}, \citenamefont {Sougrati}, \citenamefont {Jaouen},
  \citenamefont {Halevi},\ and\ \citenamefont {Mukerjee}}]{Tylus2016}%
  \BibitemOpen
  \bibfield  {author} {\bibinfo {author} {\bibfnamefont {Q.}~\bibnamefont
  {Jia}}, \bibinfo {author} {\bibfnamefont {N.}~\bibnamefont {Ramaswamy}},
  \bibinfo {author} {\bibfnamefont {U.}~\bibnamefont {Tylus}}, \bibinfo
  {author} {\bibfnamefont {K.}~\bibnamefont {Strickland}}, \bibinfo {author}
  {\bibfnamefont {J.}~\bibnamefont {Li}}, \bibinfo {author} {\bibfnamefont
  {A.}~\bibnamefont {Serov}}, \bibinfo {author} {\bibfnamefont
  {K.}~\bibnamefont {Artyushkova}}, \bibinfo {author} {\bibfnamefont
  {P.}~\bibnamefont {Atanassov}}, \bibinfo {author} {\bibfnamefont
  {J.}~\bibnamefont {Anibal}}, \bibinfo {author} {\bibfnamefont
  {C.}~\bibnamefont {Gumeci}}, \bibinfo {author} {\bibfnamefont {S.~C.}\
  \bibnamefont {Barton}}, \bibinfo {author} {\bibfnamefont {M.-T.}\
  \bibnamefont {Sougrati}}, \bibinfo {author} {\bibfnamefont {F.}~\bibnamefont
  {Jaouen}}, \bibinfo {author} {\bibfnamefont {B.}~\bibnamefont {Halevi}}, \
  and\ \bibinfo {author} {\bibfnamefont {S.}~\bibnamefont {Mukerjee}},\
  }\bibfield  {title} {\enquote {\bibinfo {title} {Spectroscopic insights into
  the nature of active sites in iron--nitrogen--carbon electrocatalysts for
  oxygen reduction in acid},}\ }\href@noop {} {\bibfield  {journal} {\bibinfo
  {journal} {Nano Energy}\ }\textbf {\bibinfo {volume} {29}},\ \bibinfo {pages}
  {65--82} (\bibinfo {year} {2016})}\BibitemShut {NoStop}%
\bibitem [{\citenamefont {Aoyama}\ \emph {et~al.}(2018)\citenamefont {Aoyama},
  \citenamefont {Kaiwa}, \citenamefont {Chantngarm}, \citenamefont
  {Tanibayashi}, \citenamefont {Saito}, \citenamefont {Hasegawa},\ and\
  \citenamefont {Nishidate}}]{Aoyama2018}%
  \BibitemOpen
  \bibfield  {author} {\bibinfo {author} {\bibfnamefont {S.}~\bibnamefont
  {Aoyama}}, \bibinfo {author} {\bibfnamefont {J.}~\bibnamefont {Kaiwa}},
  \bibinfo {author} {\bibfnamefont {P.}~\bibnamefont {Chantngarm}}, \bibinfo
  {author} {\bibfnamefont {S.}~\bibnamefont {Tanibayashi}}, \bibinfo {author}
  {\bibfnamefont {H.}~\bibnamefont {Saito}}, \bibinfo {author} {\bibfnamefont
  {M.}~\bibnamefont {Hasegawa}}, \ and\ \bibinfo {author} {\bibfnamefont
  {K.}~\bibnamefont {Nishidate}},\ }\bibfield  {title} {\enquote {\bibinfo
  {title} {Oxygen reduction reaction of fen4 center embedded in graphene and
  carbon nanotube: Density functional calculations},}\ }\href {\doibase
  10.1063/1.5053151} {\bibfield  {journal} {\bibinfo  {journal} {AIP Advances}\
  }\textbf {\bibinfo {volume} {8}},\ \bibinfo {pages} {115113} (\bibinfo {year}
  {2018})}\BibitemShut {NoStop}%
\bibitem [{\citenamefont {Allerdt}\ \emph {et~al.}(2020)\citenamefont
  {Allerdt}, \citenamefont {Hafiz}, \citenamefont {Barbiellini}, \citenamefont
  {Bansil},\ and\ \citenamefont {Feiguin}}]{Allerdt2020}%
  \BibitemOpen
  \bibfield  {author} {\bibinfo {author} {\bibfnamefont {A.}~\bibnamefont
  {Allerdt}}, \bibinfo {author} {\bibfnamefont {H.}~\bibnamefont {Hafiz}},
  \bibinfo {author} {\bibfnamefont {B.}~\bibnamefont {Barbiellini}}, \bibinfo
  {author} {\bibfnamefont {A.}~\bibnamefont {Bansil}}, \ and\ \bibinfo {author}
  {\bibfnamefont {A.~E.}\ \bibnamefont {Feiguin}},\ }\bibfield  {title}
  {\enquote {\bibinfo {title} {Many-body effects in fen4 center embedded in
  graphene},}\ }\href {\doibase 10.3390/app10072542} {\bibfield  {journal}
  {\bibinfo  {journal} {Applied Sciences}\ }\textbf {\bibinfo {volume} {10}}
  (\bibinfo {year} {2020}),\ 10.3390/app10072542}\BibitemShut {NoStop}%
\end{thebibliography}%

\clearpage

\setcounter{page}{1}
\setcounter{equation}{0}
\setcounter{figure}{0}
\setcounter{table}{0}
\renewcommand{\theequation}{S\arabic{equation}}
\renewcommand{\thetable}{S\Roman{table}}
\renewcommand{\thefigure}{S\arabic{figure}}

\onecolumngrid

\begin{center}

{\large Supplemental Material}
\vskip3mm

{\bf\large A graphene edge-mediated quantum gate}
\vskip5mm

Phillip Weinberg,$^{1}$
\vskip3mm
Adiran E. Feiguin,$^{1}$
\vskip3mm
{\noindent\it\small  
{$^1$Department of Physics, Northeastern University, Boston, Massachusetts 02115, USA}
}\vskip8mm

\end{center}

\begin{figure}[h]
	\centering
	\includegraphics[width=3in]{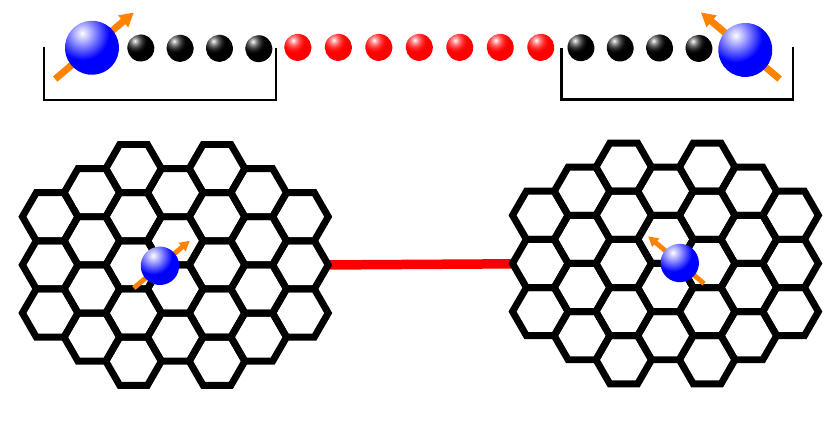}
	\caption{The top portion depicts the equivalent 1D model representing our proposed device in the calculations (see main text). The blue sphere with the arrow represents the Kondo impurity, while the black sphere represent the electronic orbitals in the flake. The red spheres in the middle of the chain is the conducting bridge between the two flakes. The bottom panel shows a possible experimental implementation. A magnetic impurity sites at the center of graphene flakes connected by a 1D conducting channel.}
	\label{sup:fig:bridge_fig}
\end{figure}

\section{Graphene Flake Mapping}

Here we discuss how we constructed the model for the quantum gate discussed on the main text. We apply a numerical approach that has been used previously to study the two impurity Kondo problems in monolayer graphene\cite{Allerdt2017}. The method is based on a technique developed in Refs.\cite{Busser2013,Allerdt.Kondo,Frontiers_review}. In those works, it was shown that single impurity Hamiltonian can be mapped, via an exact unitary transformation, onto the problem of a spin connected to a one-dimensional non-interacting chain. The hopping matrix elements of the chain are position dependent and are such that they reproduce the same physics and density of states of the original Hamiltonian: 
\begin{equation}
	H_{\rm Kondo} = J\mathbf{S}_{imp}\cdot \mathbf{s}_1 \\+ \sum_{\sigma,r=1}^\infty a_r n_{\sigma r} + \left(b_r c^\dagger_{\sigma r} c_{\sigma r+1} + {\rm h.c.}\right)
\end{equation}
Each orbital on the chain corresponds to a linear combination of single particle orbitals on the original honeycomb lattice. One can visualize these single-particle basis states as concentric ``circles'' expanding radially away from the impurity. Hence, each successive orbital in the chain corresponds to an orbital farther away from the impurity site. The orbitals close to the impurity obey the local point group symmetry of the lattice while orbitals farther away look more as circular when viewed on a macroscopic scale. One can then think of the position of the site along the chain as the ``radius'' of the single-particle wave functions in the original lattice. In practice, the chain may be truncated to some length $R$:
\begin{equation}
	H = J\mathbf{S}_{imp}\cdot \mathbf{s}_1 + \sum_{\sigma,r=1}^{R} a_r n_{\sigma r} + \sum_{\sigma,r=1}^{R-1} \left(b_r c^{\dagger}_{\sigma r} c_{\sigma r+1} + {\rm h.c.}\right)
\end{equation}
In this work we treat these truncated graphene chains of length $R$ as an approximation for an arbitrary finite-size graphene flake with a radius of $R$. This is justified when the graphene flake obeys the point group symmetry of the infinite lattice, in which case the correspondence is exact. Using these graphene chains as building blocks we can build our model for the device. We use a non-interacting tight binding chain of length $L$ with nearest neighbor hopping to represent the one dimensional channel connecting the two flakes. We will denote the chain orbitals with operators $f_{\sigma i}$ for $i=1,2,\dots,L$. For simplicity, we consider that the end of the chain will couple to the ends of each graphene chain as expressed by the full Hamiltonian: 
\begin{equation}
	H_{2K} = -t\sum_{\sigma,i=1}^{L}\left(f^\dagger_{\sigma i}f_{\sigma i+1} + {\rm h.c.}\right) + V_g\sum_{\sigma,i=1}^{L}f^\dagger_if_i+ H_1+H_2 +t'\sum_{\sigma}\left(f^\dagger_{\sigma 1} c_{\sigma R1} + f^\dagger_{\sigma L} c_{\sigma R2}+{\rm h.c.}\right) \label{eq:model}   
\end{equation}
This equivalent problem is pictorically illustrated in Fig.~\ref{fig:bridge_fig}, the black spheres on the left and right side of the chain (bracketed from below) represent the impurities and graphene flake orbitals $c_{\sigma r1}$ and $c_{\sigma r2}$ respectively. The bigger blue spheres represent the two impurities $S_1$ and $S_2$. Finally, the red spheres at the center represent the conducting bridge orbitals, $f_{\sigma i}$. As a results, the Hamiltonian representing the two flakes and the bridge system is mapped onto a 1D problem that is amenable to accurate numerical calculations, as we discuss in the main text. We point out that a realistic flake would have an irregular shape and cannot be described in terms of a single parameter $R$. The reason why we are able to do so is that the Lanczos mapping described in the text generates wave functions with the point symmetry of the lattice that grows radially away from the position of the impurity as concentric circle-like orbitals. This simplifies the formulation of the problem a great deal, but applying the mapping to a flake with an irregular shape would make the treatment more complicated. However, we would not expect the results to vary dramatically and the physics to be preserved.   

\section{Role of $t'$}

Here we briefly discuss the role of $t'$ on the correlations between the impurities. In general, for $t'\ll t$ and $t'\gg t$ leads to ground states where the impurities are free. In the limit with $t' \ll t$ the correlations will go to $0$ because the Kondo impurities will form singlets with their respective graphene flakes. In the case where $t'\gg t$ the moments are also free. In the AFM regions, the correlations will saturate similar to what happens in the $J\rightarrow 0$ limit while for the FM regions (when SP states cross $E=0$) the energy required to create a particle-hole excitation becomes too large leading to a smaller FM region which eventually disappears as $t'\rightarrow\infty$.

\end{document}